\documentclass[onecolumn,amsmath,amssymb,12pt,superscriptaddress,nofootinbib]{revtex4}
\pdfoutput=1

\usepackage[latin1]{inputenc}
\usepackage[english]{babel}
\usepackage{amssymb}
\usepackage{amsmath}
\usepackage{amsthm}
\usepackage[]{graphicx}
\usepackage[]{subfigure}
\usepackage{tensor}
\usepackage{color}
\usepackage{cancel}
\usepackage{setspace}
\usepackage{fancyhdr}
\usepackage[bookmarks,linktocpage, colorlinks=true, plainpages = false, citecolor = blue,  linkcolor=blue, urlcolor = blue, filecolor = blue]{hyperref} 

\newcommand{\smallWidthLeft}{222.5pt}
\newcommand{\smallWidthRight}{212.5pt}
\newcommand{\fullWidth}{270pt}

\begin{document}

\allowdisplaybreaks
\begin{titlepage}

\title{Classical Inflationary and Ekpyrotic Universes in the No-Boundary Wavefunction \vspace{.3in}}

\author{Jean-Luc Lehners\\ {\it Max--Planck--Institute for Gravitational Physics (Albert--Einstein--Institute), 14476 Potsdam, Germany \\ jlehners@aei.mpg.de}}


\begin{abstract}
\vspace{.3in} \noindent This paper investigates the manner in which classical universes are obtained in the no-boundary quantum state. In this context, universes can be characterised as classical (in a WKB sense) when the wavefunction is highly oscillatory, i.e. when the ratio of the change in the amplitude of the wavefunction becomes small compared to the change in the phase. In the presence of a positive or negative exponential potential, the WKB condition is satisfied in proportion to a factor $e^{-(\epsilon - 3)N/(\epsilon -1)},$ where $\epsilon$ is the (constant) slow-roll/fast-roll parameter and $N$ designates the number of e-folds. Thus classicality is reached exponentially fast in $N$, but only when $\epsilon < 1$ (inflation) or $\epsilon > 3$ (ekpyrosis). Furthermore, when the potential switches off and the ekpyrotic phase goes over into a phase of kinetic domination, the level of classicality obtained up to that point is preserved. Similar results are obtained in a cyclic potential, where a dark energy plateau is added. Finally, for a potential of the form $-\phi^n$ (with $n=4,6,8$), where the classical solution becomes increasingly kinetic-dominated, there is an initial burst of classicalisation which then quickly levels off. These results demonstrate that inflation and ekpyrosis, which are the only dynamical mechanisms known for smoothing the universe, share the perhaps even more fundamental property of rendering space and time classical in the first place. 
\end{abstract}
\maketitle

\end{titlepage}

\tableofcontents

\section{Introduction}

In analysing models of the early universe, the usual procedure is to consider a classical background spacetime and to calculate what happens to small quantum fluctuations around this background. One then calculates the amplitude and correlation functions of the fluctuations in order to determine whether a given model is in agreement with current cosmological observations. However, as is well known yet seldom addressed, this procedure hinges on a very large number of assumptions. Is the modelling of matter as a scalar field justified? Is it permissible to consider a particular classical background solution, given that the same theory contains many other solutions? Why is it ok to consider a split into a classical background with superimposed quantum fluctuations? In other words, why can the background be treated classically in the first place, given that the laws of nature are quantum laws? 

String theory provides a framework in which one can see how an effective scalar field driving the dynamics of the early universe may arise. In this framework, scalar fields show up in many guises, and can for instance parameterise the shape of the internal dimensions, thus providing a geometric picture for cosmological evolution. However, in this framework it can be expected that the potential for the scalar(s) is rather intricate, as there are many moving parts in the theory. In such a potential ``landscape'' many different kinds of solutions arise in different regions of the potential, adding the question of where on the potential cosmological evolution may be expected to start. 

In order to address many of these questions, a theory of initial conditions seems to be required. String theory by itself, in its current understanding, does not provide one. Nor do the canonical approaches to quantum gravity. In fact, despite their obvious relevance, not many theories of initial conditions have been developed in any detail, and all proposals have important open questions attached to them. In the present paper, we will consider the Hartle-Hawking no-boundary proposal \cite{Hawking:1981gb,Hartle:1983ai,Hawking:1983hj}, which is arguably the most attractive theory to date. The no-boundary proposal may only make sense in the semi-classical approximation to quantum gravity, and it seems likely that it will not be the final word on this topic \cite{Kiefer:1990ms,Kiefer:1992ci}. Nevertheless, the configurations we will be most interested in all involve small spacetime curvatures, and hence one may certainly hope that there are valuable insights to be obtained from this approach.

Within this approach, and more generally in studies of quantum cosmology, it was assumed until recently that inflation is necessary in order to obtain a classical spacetime \cite{Hartle:2007gi,Hartle:2008ng}. That is, it has been said that an inflationary phase is not only required to explain the data seen in the cosmic microwave background, but that one needs inflation in order to be able to talk about spacetime at all. In the present paper, we will reinforce this view by showing just how efficient inflation is in rendering spacetime classical, for the particular case when the slow-roll parameter is constant. However, the story has become more involved with the discovery of {\it ekpyrotic instantons} \cite{Battarra:2014xoa}, which are solutions obeying the no-boundary conditions and mediating the emergence of a classical ekpyrotic contracting universe. These solutions can also explain the classicality of spacetime, and moreover they have a high relative probability associated with them. We will show in the present paper that for a single scalar field and constant equation of state, inflation and ekpyrosis are in fact the only two mechanisms that can make spacetime classical. Moreover, both are similarly efficient at achieving this. 

We also analyse what happens to the classicality of spacetime after the ekpyrotic phase comes to an end. When the potential turns off, the universe becomes dominated by the kinetic energy of the scalar field, and as we will show, the level of classicality achieved up to that moment is preserved during the kinetic phase. Thus, the universe remains highly classical as it approaches the bounce. We should point out that we have not included a description of the bounce. An important open question is therefore whether a bounce into an expanding phase of the universe can be successfully incorporated -- this question is left for follow-up work.

Our framework further allows us to study how classical spacetime emerges from a cyclic potential, where a dark energy plateau is added to the ekpyrotic phase. One may rightfully ask why we need to address the issue of initial conditions in a cyclic universe, given that each cycle sets up the ``initial'' conditions for the next one. The reason is that in the cyclic universe, in order to avoid the build-up of entropy, each cycle must grow larger than the previous one. Thus, any finite region of the universe (or the whole universe, if it is finite) was sub-Planckian a finite time into the past, and requires initial conditions. This suggests the view that a cyclic universe alleviates the issue of initial conditions, as it may allow them to become progressively fine-tuned in a dynamical fashion from cycle to cycle, but it cannot avoid the issue entirely, as the emergence of the first cycle remains to be explained. For illustration, we will compare two different histories (leading up to the first bounce), one in which the universe is always contracting and one where the universe is first expanding (due to the dark energy plateau in the potential) and then contacting. We will find that the second history is both more likely and achieves a greater amount of classicality. Finally, we will look at steep negative potentials with power-law form $V=-\phi^n$ for $n=(4,6,8).$ These potentials lead to a kind of ``transient'' ekpyrotic phase, where the equation of state is very large initially but quickly drops down to that of a kinetic-dominated universe. Correspondingly, we find that for such potentials there is a brief burst of classicalisation, which comes to a halt as the universe becomes kinetic-dominated. As in the case described above, the level of classicality achieved is then maintained during kinetic domination.

Our results thus may help in answering some of the questions posed at the beginning of this introductory section.
 
\section{Quantum Cosmology and WKB Classicality} \label{section:QC}

To begin, we will review/discuss a number of standard results in quantum cosmology. We will consider theories of gravity minimally coupled to a scalar field $\phi$ with a potential $V(\phi),$ with action
\begin{eqnarray}
S &=& \int d ^4 x  \sqrt{-g} \left( \frac{R}{2 \kappa^2} - \frac{1}{2} g ^{\mu \nu} \partial _{\mu} \phi\, \partial _{\nu} \phi - V( \phi) \right) \;. \label{LorentzianAction}
\end{eqnarray}
Moreover, we will restrict our analysis to the simplified context of minisuperspace, i.e. we will restrict to closed Friedmann-Lemaitre-Roberston-Walker universes with metric
\begin{eqnarray} \label{eq:LorentzianMetric}
ds ^2 & = & - \tilde{N} ^2( \lambda) d \lambda ^2 + a ^2( \lambda) d \Omega _3 ^2 \;,
\end{eqnarray}
where $\tilde{N}$ is the lapse function and $d\Omega_3^2$ the metric on the unit three-sphere, with the scalar field also depending solely on time $\phi=\phi(\lambda).$ The action then becomes
\begin{eqnarray}
S &=& \frac{6 \pi ^2}{ \kappa^2} \int d \lambda \tilde{N} \left( - a \frac{ \dot{a} ^2 }{\tilde{N} ^2} + a + \frac{ \kappa^2 a ^3}{3} \left( \frac{1}{2} \frac{ \dot{ \phi} ^2}{\tilde{N} ^2} - V \right) \right) \;,
\end{eqnarray}
where $\; \dot{} \equiv d/d \lambda$. Following \cite{Hartle:2008ng}, we will re-write the action in the form\footnote{For simplicity, we re-scale the action by $6\pi^2$ here. This factor is re-introduced from Eq. \eqref{EuclidAction} onwards.}
\begin{eqnarray}
S &=& \int d\lambda \tilde{N} \left( \frac{1}{2} G_{AB} \frac{1}{\tilde{N}}\frac{dq^A}{d\lambda}\frac{1}{\tilde{N}}\frac{dq^B}{d\lambda} - U(q^A) \right),
\end{eqnarray}
with $q^A = (a,\phi)$ and $G_{aa} = -2a, \, G_{\phi\phi} = \kappa^2 a^3/3.$ Then we have an associated Hamiltonian
\begin{eqnarray}
{\cal H} &=& \frac{1}{2} G^{AB} p_A p_B + U \, , \label{Hamiltonian}
\end{eqnarray}
with the canonical momenta $p_a = -2a \dot{a}, \, p_\phi = \frac{\kappa^2 a^3}{3} \dot\phi,$ and where the effective potential is given by
\begin{eqnarray}
U &=& -a + \frac{\kappa^2 a^3}{3} V \, .
\end{eqnarray} 
We have set the lapse function $\tilde{N}$ to unity and from now on we will also set the gravitational coupling $\kappa$ to unity. The Hamiltonian is classically zero and corresponds to the Friedmann equation. If one quantises the theory canonically, by replacing $p_A \rightarrow -i\hbar \frac{\partial}{\partial q^A},$ one obtains the quantum version of the Hamiltonian constraint, namely the Wheeler-deWitt (WdW) equation
\begin{eqnarray}
\hat{\cal H} \Psi &=& \left( - \frac{\hbar^2}{2} G^{AB} \frac{\partial}{\partial q^A}\frac{\partial}{\partial q^B} + U \right) \Psi = 0\, ,
\end{eqnarray}
where $\Psi = \Psi(a,\phi)$ is the wavefunction of the universe. In our case the WdW equation reads
\begin{equation}
\left( \frac{\hbar^2}{4a}\frac{\partial^2}{\partial a^2} - \frac{3\hbar^2}{2a^3} \frac{\partial^2}{\partial \phi^2} - a + \frac{a^3}{3} V \right) \Psi = 0 \, \label{WdW}
\end{equation}
where we have ignored factor ordering ambiguities in the first term (since these are unimportant for the purposes of the present discussion). The question now is: under what conditions can the wavefunction be interpreted as corresponding to a classical universe?

We can make progress on this issue by writing the wavefunction in the form
\begin{equation}
\Psi =  e^{(- A + i S)/\hbar}, \label{wfAnsatz}
\end{equation}
where $A(a,\phi), S(a,\phi)$ are real functions. We can then expand the WdW equation in powers of $\hbar.$ The real and imaginary parts of the two leading orders yield
\begin{eqnarray}
\frac{1}{2} (\nabla A)^2-\frac{1}{2} (\nabla S)^2 + U &=& 0, \qquad \nabla A \cdot \nabla S = 0, \label{WdW1} \\ \nabla^2 A &=& 0, \qquad \nabla^2 S = 0, \label{WdW2}
\end{eqnarray}
where we have defined $\nabla^2 \equiv G^{AB} \partial_A \partial_B $. If $|\partial_A A| \ll |\partial_A S|,$ i.e. if the amplitude of the wavefunction is slowly varying compared to its phase, then from the first equation above it follows that $S$ approximately satisfies the Hamilton-Jacobi equation,
\begin{equation}
-\frac{1}{2} (\nabla S)^2 + U \approx 0. 
\end{equation} 
This corresponds to the Wentzel-Kramers-Brillouin (WKB) semi-classical approximation. In this case, the wavefunction is strongly peaked along classical solutions of the equations of motion, which are characterised by the first integral  (cf. Eq. \eqref{Hamiltonian})
\begin{equation}
p_A = \frac{\partial S}{\partial q^A}. \label{FirstIntegral}
\end{equation}  
It is in this sense that oscillating wave functions are associated to an ensemble of classical solutions and thus to classical spacetime and matter configurations\footnote{A canonical transformation from $(p=\frac{\partial G}{\partial q},q)$ to the new variables $(\bar{p},\bar{q} = \frac{\partial G}{\partial \bar{p}}),$ using the generating function $G(q,\bar{p}),$ leads to a new wavefunction $\bar{\Psi}(\bar{p}) = \int dq e^{-iG} \Psi(q).$ Using the generating function $G=q\bar{p} + S(q),$ the wavefunction $\Psi=e^{iS(q)}$ then gets transformed into $\bar\Psi = \delta(\bar{p})$ where $\bar{p} = p - \frac{\partial S}{\partial q},$ demonstrating that in the WKB approximation the wavefunction is strongly peaked along configurations characterised by $p=\frac{\partial S}{\partial q}$  \cite{Halliwell:1990uy}.}.  

In addition, the second equation in \eqref{WdW1} implies that along such classical trajectories the amplitude $A$ is (approximately) conserved. Together with \eqref{WdW2} this further implies the relation  
\begin{equation}
\nabla \cdot \left( e^{-2 A} \nabla S \right) = 0,
\end{equation}
which can also be obtained as a consequence of the conservation (via the WdW equation) of the Klein-Gordon current
\begin{equation}
J = \frac{i}{2} (\Psi^\star \nabla \Psi - \Psi \nabla \Psi^\star).
\end{equation} 
This suggests that when the WKB approximation holds, we may choose surfaces where the normal derivative $\nabla S > 0$ and use $e^{-2A}$ as a measure of the relative probability of a given classical history (see e.g. \cite{Vilenkin:1988yd,Halliwell:1990uy} for more details).
 
The above arguments are somewhat heuristic, but they provide a simple and coherent framework for interpreting the wavefunction in the WKB approximation. For attempts to make all of this more precise using the decoherent histories approach, see e.g. \cite{Halliwell:2011zz}. Note also that the notion of classicality we have discussed so far only applies to a single component of the wavefunction. In general the semi-classical approximation will yield a wavefunction composed of a sum of terms of the form \eqref{wfAnsatz}. Then one must also determine under what conditions the different components do not interfere with one another -- in other words, the different components must decohere. We will not analyse this issue in detail, but offer additional comments in the discussion section.

Given the above discussion of WKB classicality, we may wonder under what conditions oscillatory wave functions may be expected. Inspection of the WdW equation \eqref{WdW} allows us to single out the following  cases:
\begin{itemize}
\item Inflation: $V>0.$ \\ If the field is slowly rolling and the universe is rapidly expanding, we can assume $\frac{1}{a^2}\frac{\partial^2}{\partial \phi^2} \ll \frac{\partial^2}{\partial a^2}$. Then at large scale factor $a$ we are left with the equation
\begin{equation}
\left( \frac{\hbar^2}{4a}\frac{\partial^2}{\partial a^2}  + \frac{a^3}{3} V \right) \Psi = 0 \, 
\end{equation}
and thus we can expect an oscillatory solution for $\Psi.$
\item Ekpyrosis: $V<0.$ \\ If the field is fast-rolling and the universe slowly contracting, we have the opposite situation, namely that $\frac{1}{a^2}\frac{\partial^2}{\partial \phi^2} \gg \frac{\partial^2}{\partial a^2}$. Then we can ignore the first term in the WdW equation, leading to the equation
\begin{equation}
\left(\frac{3\hbar^2}{2a^3} \frac{\partial^2}{\partial \phi^2} + a - \frac{a^3}{3} V \right) \Psi = 0 
\end{equation}
and once again to an oscillating solution for $\Psi$ since the potential is negative.
\end{itemize}
The aim of the present paper is to study in detail how (and how fast) this oscillatory behaviour of the wavefunction is reached in these two regimes.

So far, we have not said anything yet about boundary conditions. For this reason, the wave functions under discussion were peaked on an ensemble of classical solutions. If we want to specify which of these solutions is the relevant one, we must impose boundary conditions. A particularly compelling way of doing this is the Hartle-Hawking ``no-boundary'' proposal \cite{Hawking:1981gb,Hartle:1983ai,Hawking:1983hj}, which is formulated in the path integral approach to the problem. Indeed, the wavefunction can also be calculated using a path integral,
\begin{equation}
\Psi = \int_{\cal C} \delta g\, \delta \phi \, e^{-S_E(g_{\mu\nu},\phi)}  \;.
\end{equation} 
Imposing boundary conditions is then equivalent to restricting the class of paths ${\cal C}$ one is summing over in the path integral. Above we have already performed a Wick rotation to the Euclidean action $S_E,$ which in the minisuperspace approximation is given by 
\begin{equation}
S_E = 6 \pi ^2 \int d \lambda N \left( - a \frac{ \dot{a} ^2 }{N ^2} - a + \frac{a ^3}{3} \left( \frac{1}{2} \frac{ \dot{ \phi} ^2}{N ^2} + V \right) \right) \;, \label{EuclidAction}
\end{equation} 
where now $N=i \tilde{N}.$ Allowing complex functions of $ \lambda$ (as will be necessary), the integral can be interpreted as a contour integral in the complex plane, with $d\tau \equiv N d \lambda$,
\begin{equation} \label{eq:complexAction}
S_E = 6 \pi ^2 \int d \tau \left( - a a ^{\prime 2} - a + \frac{a ^3}{3} \left( \frac{1}{2} \phi ^{\prime 2} + V \right) \right) \;,
\end{equation}
where $\; ^{\prime} \equiv d/d \tau$. The constraint and equations of motion following from this action are
\begin{eqnarray}
a ^{\prime 2}  -1  - \frac{a ^2}{3} \left( \frac{1}{2} \phi ^{\prime 2} - V \right) &=& 0 \;, \\
\phi '' + 3 \frac{ a'}{a} \phi' - V_{, \phi} & = & 0 \;,\\
a'' + \frac{ \kappa^2 a}{3} \left( \phi ^{\prime 2} + V \right) & = & 0 \;.
\end{eqnarray}
Using the constraint, the on--shell action then simplifies to
\begin{equation} \label{eq:onshellAction}
S_E ^{on-shell} =  4 \pi ^2 \int d \tau\, \left(- 3 a + a ^3 V \right) \;.
\end{equation}

\begin{figure}[h]
\centering
\begin{minipage}{\fullWidth}
\includegraphics[width=\fullWidth]{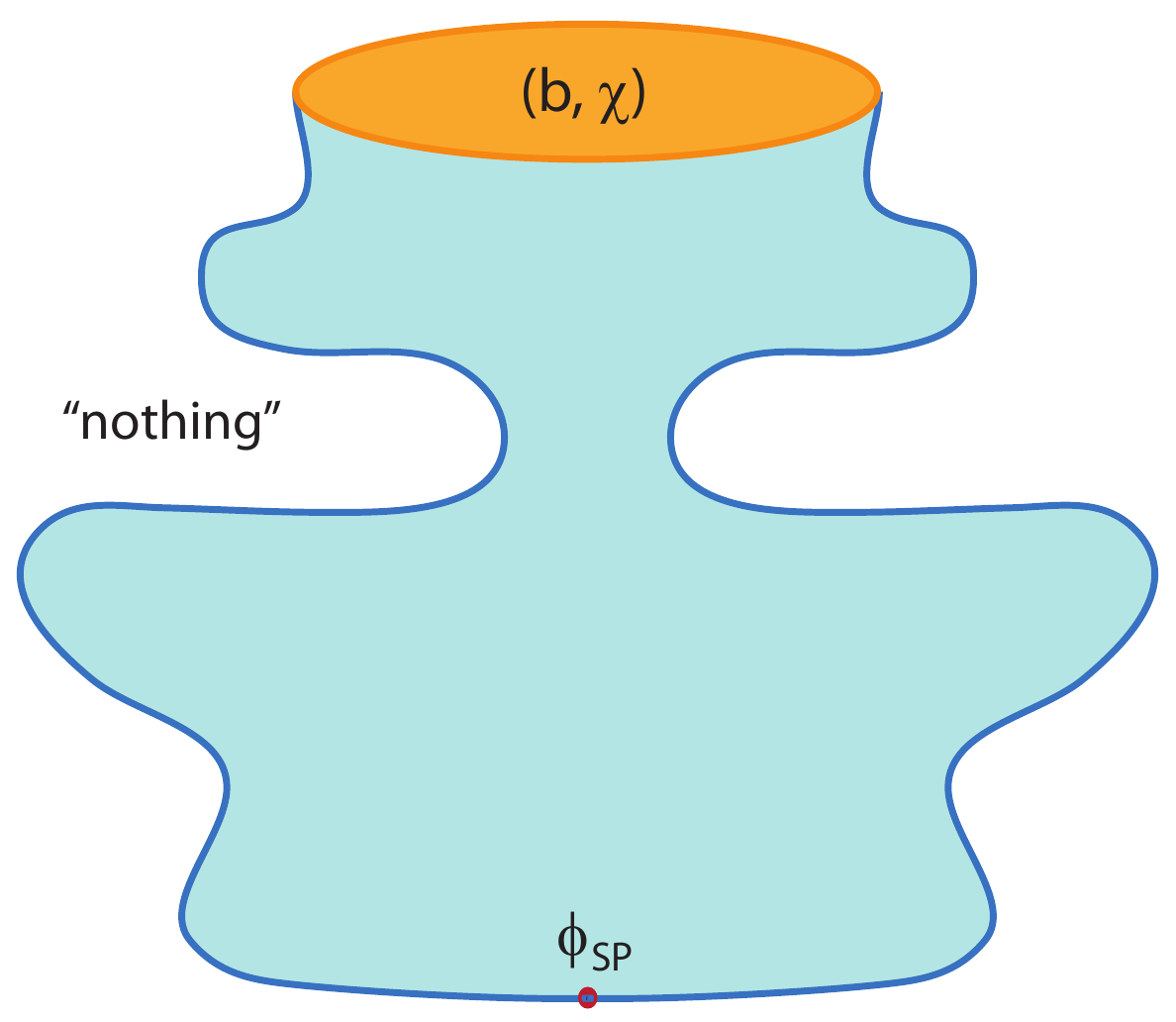}
\end{minipage}%
\caption{\label{fig:NoBoundary} In the no-boundary proposal, the path integral for the wavefunction $\Psi(b,\chi)$ is summed over geometries that contain no boundary to the past, and that take the specified values $a(\tau_f) = b, \phi(\tau_f) = \chi$ on a final hypersurface (in orange).}
\end{figure}

We are now in a position to specify the no-boundary proposal. In the minisuperspace approximation, the path integral reduces to the form 
\begin{equation}
\Psi( b, \chi) = \int_{\cal{C}} \delta a \delta \phi \, e^{-S_E(a,\phi)}  \;,
\end{equation}
where the arguments $b$ and $\chi$ are the values of the scale factor and scalar field at the time of interest. The integral is performed only over paths (4-geometries) $\cal{C}$ which have no boundary in the past and for which the scale factor and scalar field take the (real) specified values $b$ respectively $\chi$ on a final hypersurface -- see Fig. \ref{fig:NoBoundary}. The central idea is that in this way the universe is entirely self-contained, in both space {\it and} time. In practice, we will evaluate the path integral in the saddle point approximation,
\begin{equation}
\Psi( b, \chi)  \sim \sum e^{- S_E(b, \chi)} \;,
\end{equation}
where $S_E(b, \chi)$ denotes the Euclidean action of a {\it complex} instanton solution $(a(\tau), \phi( \tau))$ of the equations of motion. The no-boundary proposal then translates into the following boundary conditions for the instantons:
\begin{itemize}
\item At $a(0) = 0$ the solution must be regular. (Note that we have arbitrarily put the ``bottom'' of the instanton at $\tau=0.$) Inspection of the equations of motion shows that this can be achieved if  $a'(0)=1$ and $\phi'(0)=0.$ Hence, the instantons of interest can be labelled by the (complex) value $\phi_{SP}$ of the scalar field at the bottom, or ``South Pole'', of the instanton. 
\item At a certain final time $ \tau_f$ the scale factor and scalar field must take the real values $a(\tau_f)=b, \phi(\tau_f)=\chi.$ 
\end{itemize}

If we now want to study the properties of a given (classical) history $\left( a(\lambda), \phi(\lambda) \right)$ of the universe, then we must evaluate the wavefunction $\Psi(b,\chi)$ for successive values of $\left(b = a(\lambda),\chi= \phi(\lambda)\right)$ along this history. Moreover we need the derivatives of $\Psi$ with respect to $b$ and $\chi$ in order to determine the WKB conditions, since these are given by
\begin{eqnarray}
| \partial _{b} S_E ^{R}| & \ll & | \partial _{b} S_E ^{I}| \;,\label{WKBb} \\
| \partial _{ \chi} S_E ^{R}| & \ll & | \partial _{ \chi} S_E ^{I}| \;. \label{WKBchi}
\end{eqnarray}
where $S_E^R, S_E^I$ denote the real and imaginary components of the Euclidean action $S_E$ of the instantons. According to the discussion above, we can say that the universe becomes increasingly classical when the inequalities above are increasingly well satisfied. Under these circumstances, the real part of the action approaches a constant, and the particular history under consideration can be assigned a relative probability proportional to $e^{-2 S_E^R}.$ On the other hand, when the WKB conditions \eqref{WKBb} and \eqref{WKBchi} are not satisfied, then the wavefunction does not imply any classical correlations between the fields and their momenta, and moreover no meaningful notion of probability can be defined.

\section{Classical Spacetime from Inflation} \label{section:inflation}

\begin{figure}[h]
\centering
\begin{minipage}{\fullWidth}
\includegraphics[width=\fullWidth]{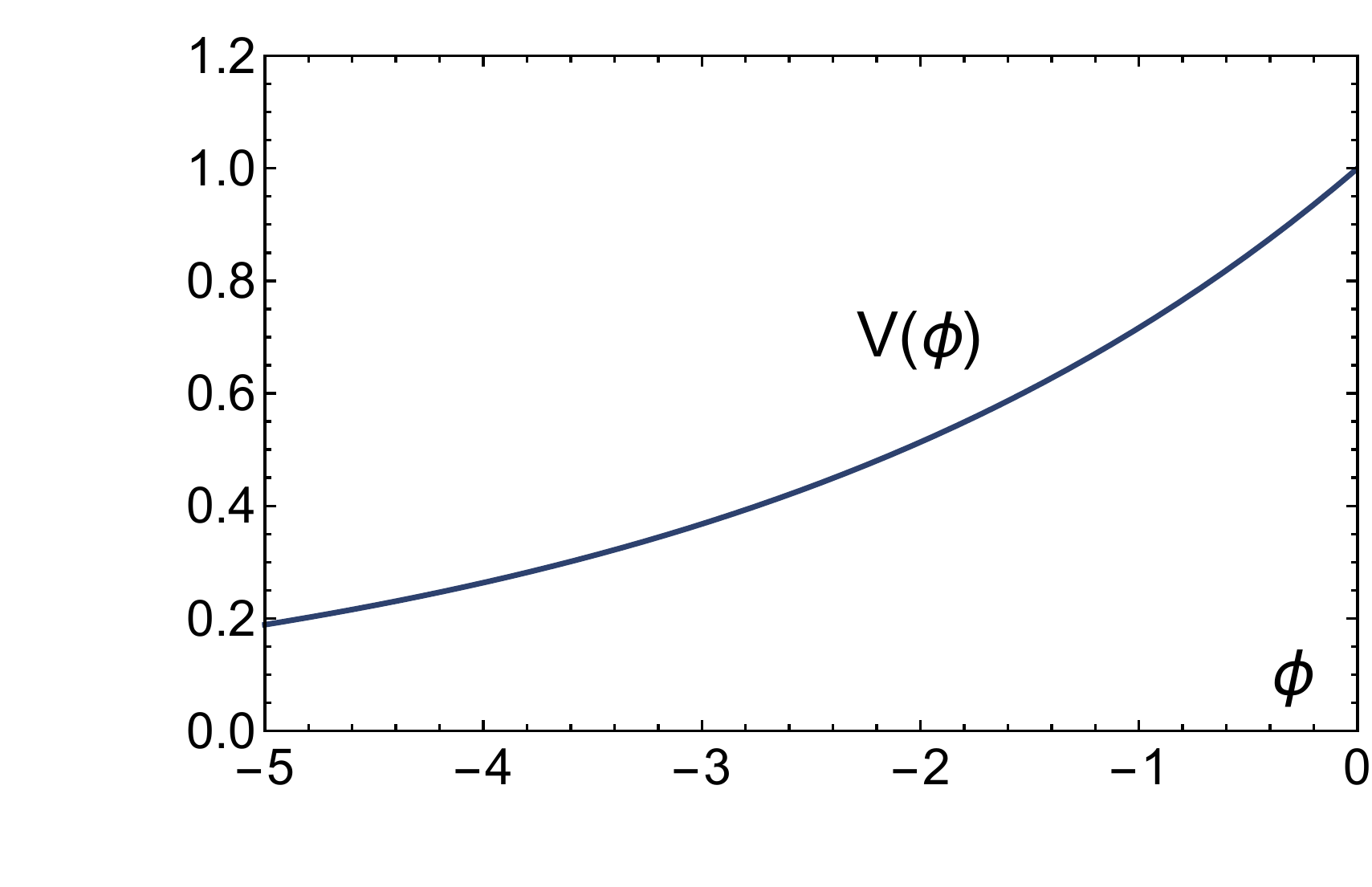}
\end{minipage}%
\caption{\label{fig:InfPot} The inflationary potential used for the numerical evaluations is given by $V(\phi) = e^{\phi/3}$. This corresponds to a slow-roll parameter $\epsilon = 1/18.$}
\end{figure}

We start by considering positive exponential potentials
\begin{equation}
V(\phi) = V_0 e^{c \phi},
\end{equation}
where $V_0> 0$ and $c$ are constants - see Fig. \ref{fig:InfPot}.
For such potentials the slow-roll parameter
\begin{equation}
\epsilon = \frac{V_{,\phi}^2}{2 V^2} = \frac{c^2}{2}
\end{equation}
is constant, and the theory admits the asymptotic scaling solution
\begin{equation}
a=a_0 (\lambda)^{1/\epsilon}, \qquad \phi = - \sqrt{\frac{2}{\epsilon}} \ln \left( \sqrt{\frac{\epsilon^2 V_0}{3-\epsilon}} \, \lambda\right), \qquad V = \frac{3-\epsilon}{\epsilon^2\lambda^2}\,. \label{scalingsolutionInf}
\end{equation} 
When $\epsilon < 1,$ this solution corresponds to accelerated expansion, i.e. inflation.

\begin{figure}[htbp]
\begin{minipage}{\smallWidthLeft}
\includegraphics[width=\smallWidthRight]{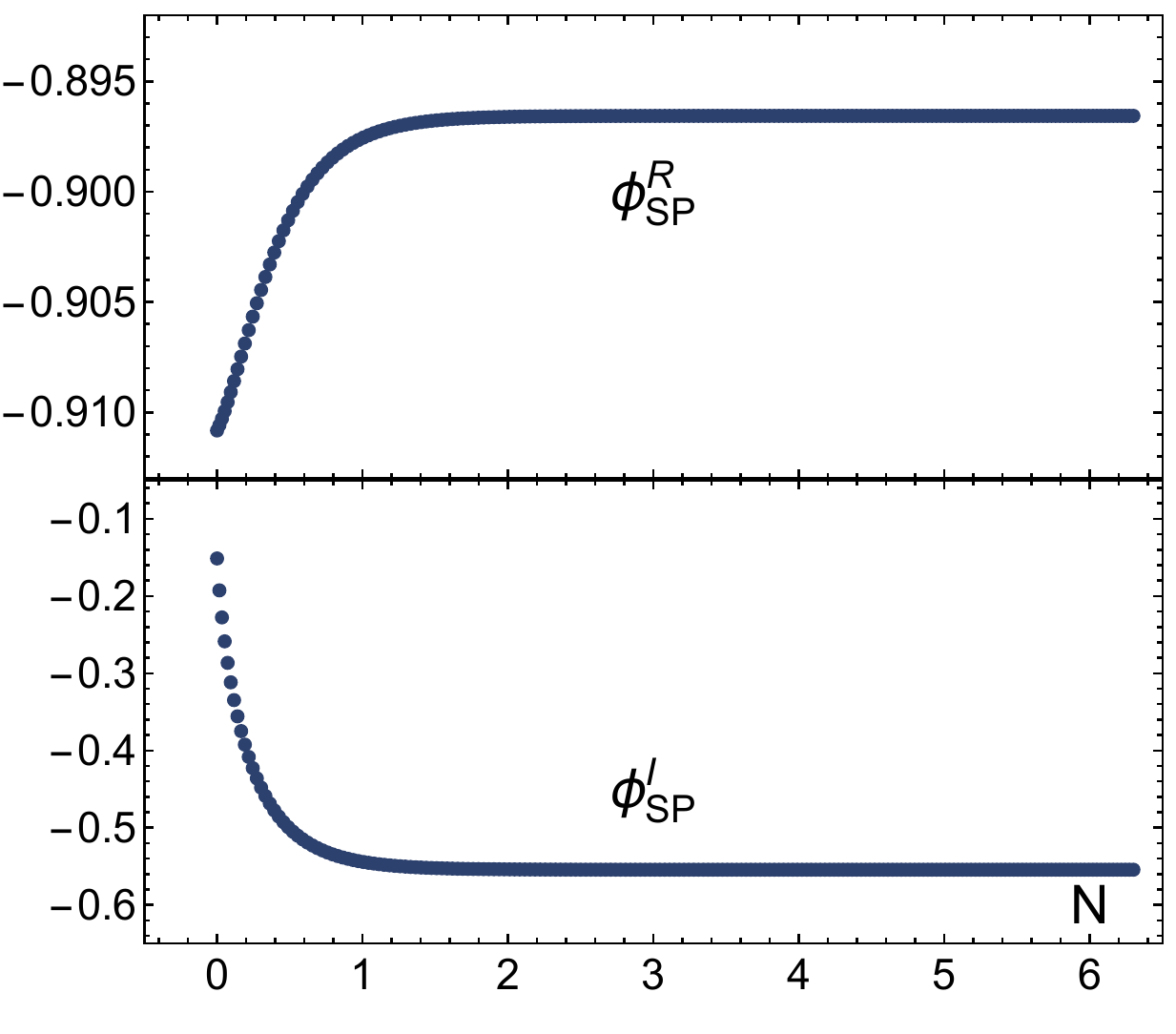}
\end{minipage}%
\begin{minipage}{\smallWidthRight}
\includegraphics[width=\smallWidthRight]{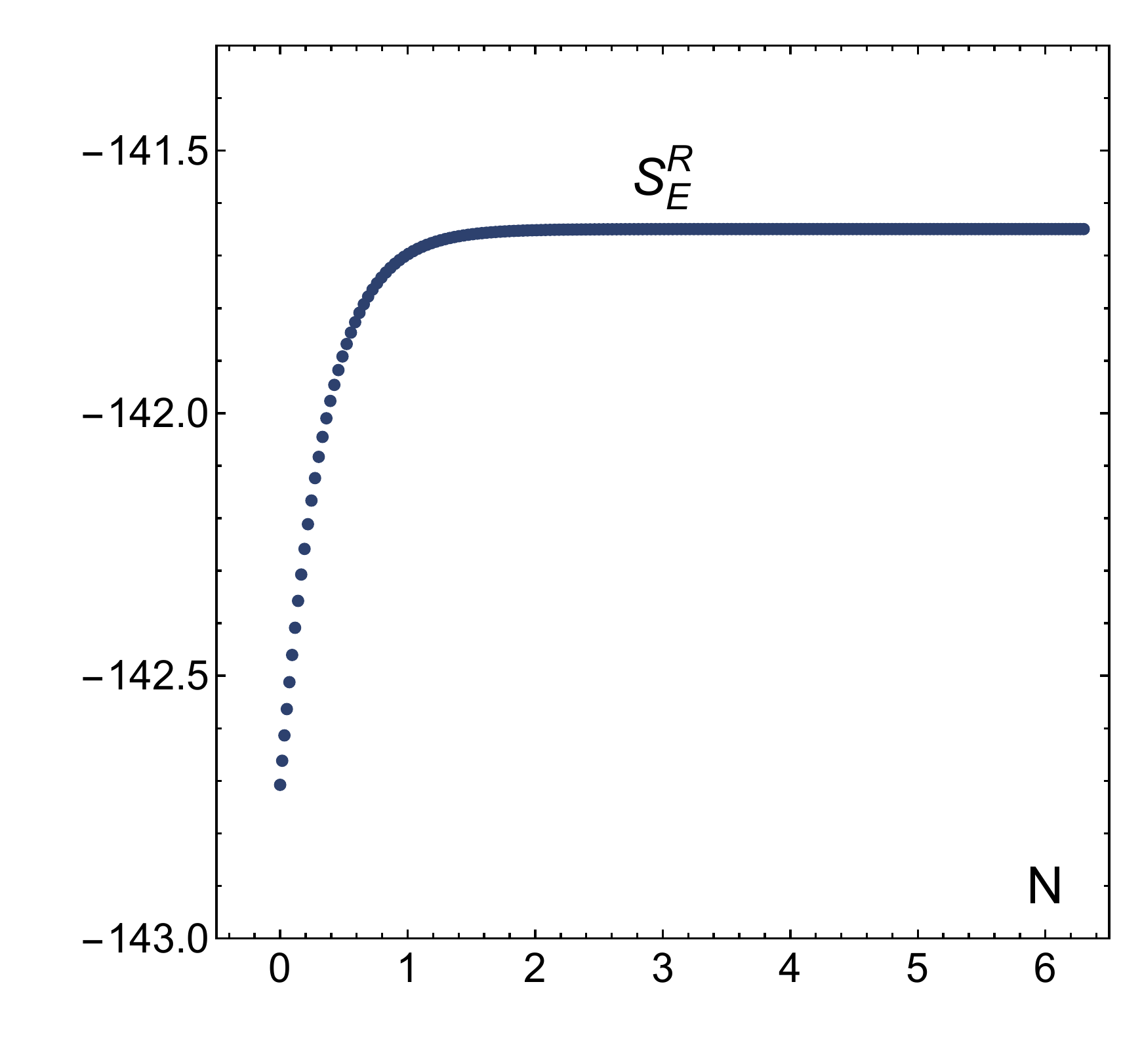}
\end{minipage}%
\caption{ \label{fig:InfPhi0ReS} {\it Left panel:} The South Pole scalar field value $\phi_{SP}=\phi_{SP}^R + i \phi_{SP}^I$ for the series of instantons corresponding to the classical history \eqref{InfIC1} -- \eqref{InfIC2}, as a function of the number of e-folds $N$ of inflation. {\it Right panel:} An analogous plot for the real part of the Euclidean action $S_E^R$ reached during the course of the same classical history.} 
\end{figure}

We are now interested in whether the wavefunction corresponding to a classical solution becomes ``classical'' in the WKB sense. As an example, we have chosen the classical solution specified by the following initial data at time $\lambda_i$:
\begin{eqnarray}
c & = \frac{1}{3}, \qquad \epsilon & = \frac{1}{18} \\ a(\lambda_i)  & = 2, \qquad \dot{a}(\lambda_i) & = \left( -1 + \frac{a(\lambda_i)^2}{3}(\frac{1}{2} \dot\phi(\lambda_i)^2 + V(\lambda_i))\right)^{1/2} \label{InfIC1} \\
\phi(\lambda_i) = & -\frac{1}{2} \qquad \dot{\phi}(\lambda_i) & = - \left( \frac{2\epsilon V(\lambda_i)}{3-\epsilon} \right)^{1/2}\, . \label{InfIC2}
\end{eqnarray}
We evaluated the wavefunction for $200$ values of $(b,\chi)$ between the initial values above and the final values $(b\approx 1578, \chi \approx -3.047).$ This corresponds to a range of a little over $6$ e-folds $N,$ where $dN=d\ln(aH)$ with $H=\dot{a}/a.$ The complex scalar field values $\phi_{SP}$ as well as the real part of the action $S_E^R$ corresponding to this classical history are shown in Fig. \ref{fig:InfPhi0ReS}. One can see very clearly that as the classical history progresses, i.e. as the universe expands during the inflationary phase, the values of $\phi_{SP}$ and $S_E^R$ stabilise, and an unambiguous relative probability of $e^{-2S_E^R}$ can be defined for this history.

\begin{figure}[htbp]
\begin{minipage}{\smallWidthLeft}
\includegraphics[width=\smallWidthRight]{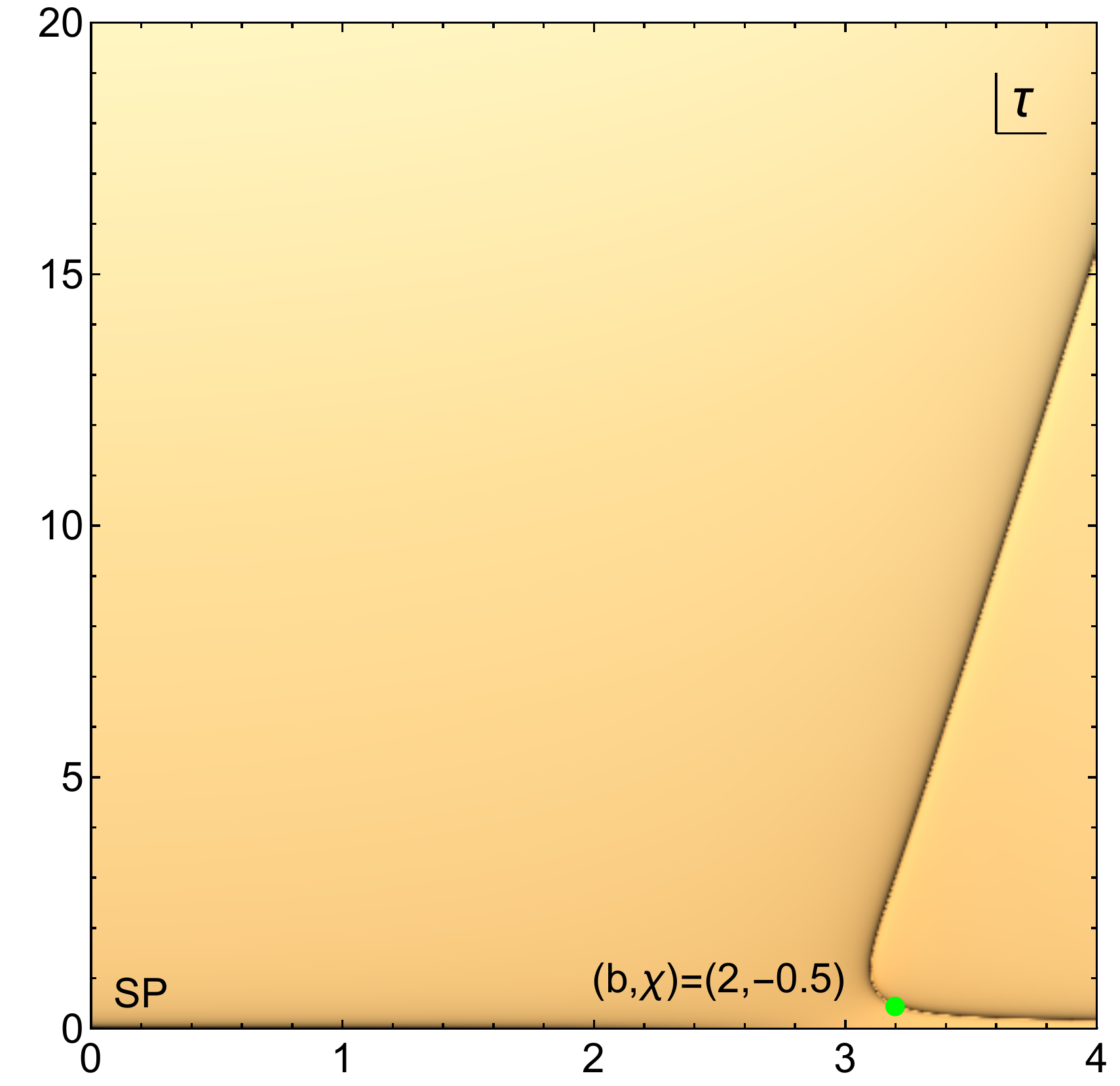}
\end{minipage}%
\begin{minipage}{\smallWidthRight}
\includegraphics[width=\smallWidthRight]{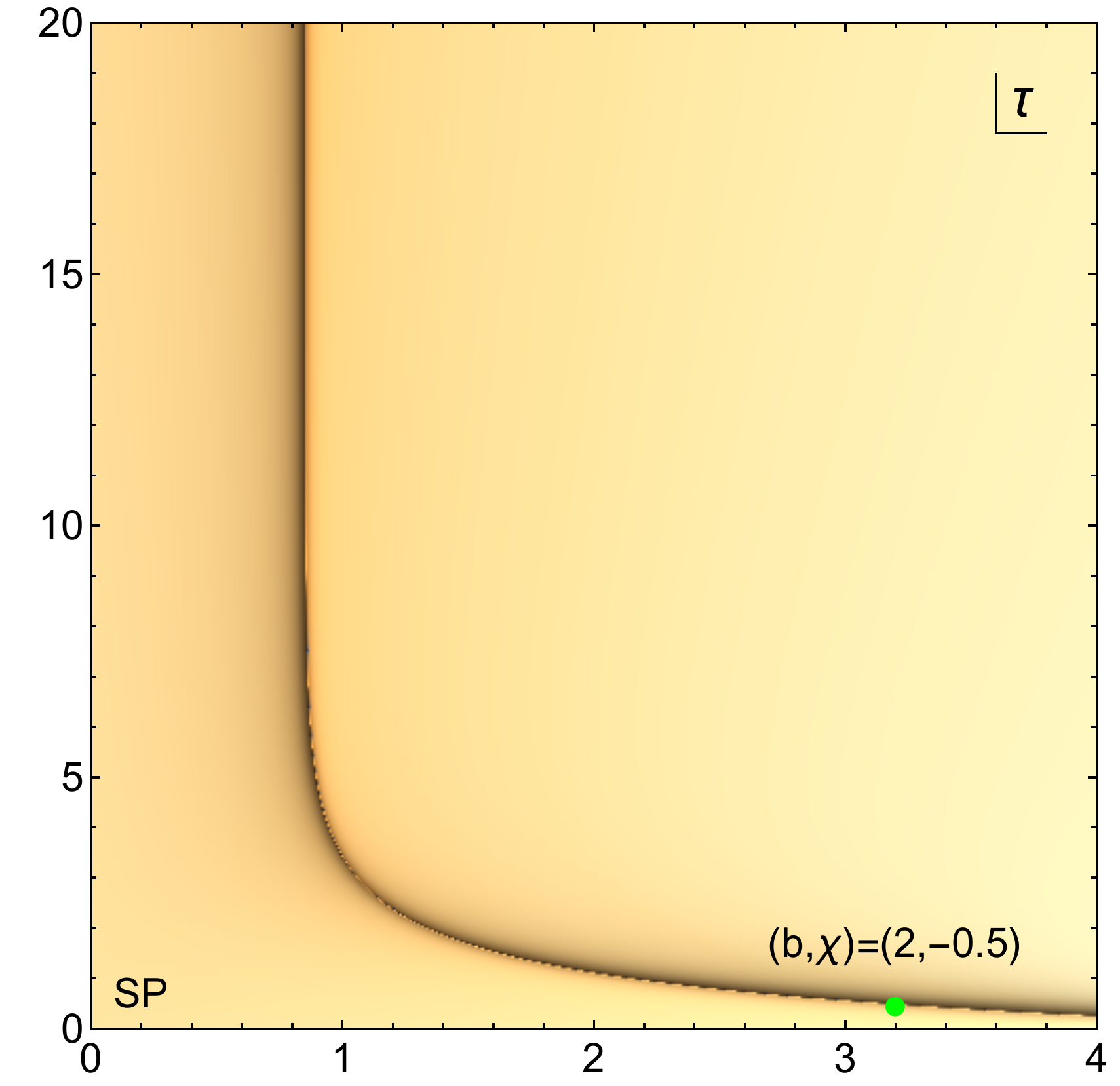}
\end{minipage}%
\caption{ \label{fig:InfEarly} A visual representation of the instanton corresponding to the values $(b=2,\chi=-1/2)$. This instanton is characterised by the value $\phi_{SP}=-0.9108-0.1513 i.$ To obtain these figures, the equations of motion are integrated from the South Pole in a dense grid. The dark lines indicate the locus of real scale factor (left panel) and real scalar field (right panel). The argument $(2,-1/2)$ of the wavefunction is reached at the green dot. Note that the wavefunction is still far from classical at this stage, as is clear from the fact that the lines of real scale factor and scalar field only overlap at a single point.} 
\end{figure}

\begin{figure}[htbp]
\begin{minipage}{\smallWidthLeft}
\includegraphics[width=\smallWidthRight]{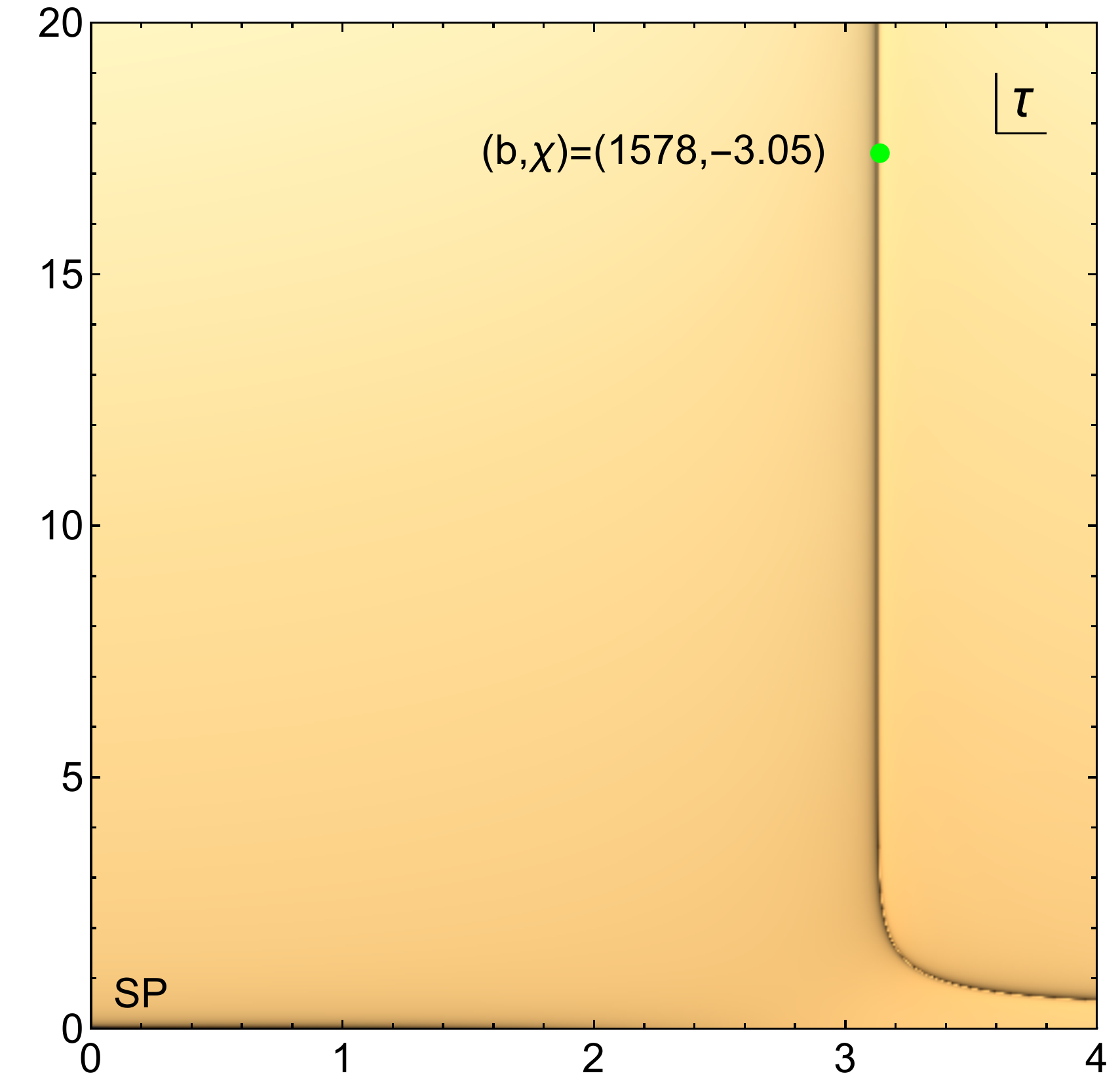}
\end{minipage}%
\begin{minipage}{\smallWidthRight}
\includegraphics[width=\smallWidthRight]{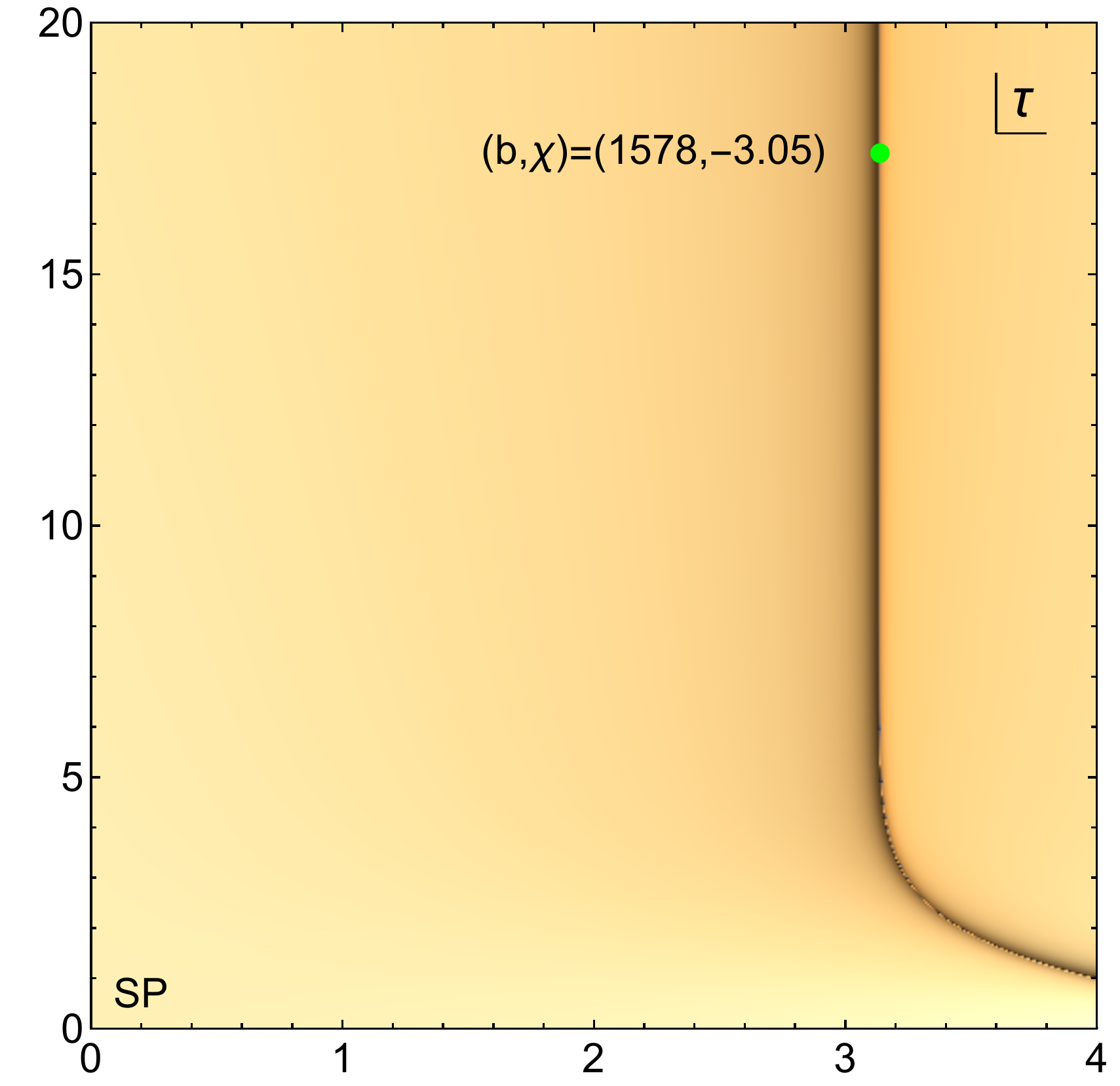}
\end{minipage}%
\caption{ \label{fig:InfLate} A visual representation of the instanton corresponding to the values $(b=1578,\chi=-3.05)$. This instanton is characterised by the value $\phi_{SP}=-0.8966-0.5544 i.$ The wavefunction is now already highly classical.} 
\end{figure}

We have illustrated the first and last instanton of the series in Figs. \ref{fig:InfEarly} and \ref{fig:InfLate} respectively. These plots have been obtained by solving the field equations from $\tau=0$ (with the corresponding value of $\phi_{SP}$) up along the imaginary axis and then horizontally across, in a dense grid so as to cover the shown region of the complex $\tau$ plane. The plots are relief plots of $Im(a)$ (left panel) and $Im(\phi)$ (right panel), with darker colours corresponding to a smaller imaginary part of the scale factor (left panel) and scalar field (right panel). Thus the black lines are the locus where the scale factor respectively the scalar field take on real values. For a classical history, these lines must be vertical and overlap. Note that for the early instanton, we are still far from classicality. The lines of real $a$ and $\phi$ cross each other at the desired values, i.e. at the green dot where the specified arguments of the wavefunction $(b=2, \chi=-1/2)$ have been reached, but elsewhere at best one of $a$ or $\phi$ is real. However, the situation changes when we look at a later instanton, such as the one depicted in Fig. \ref{fig:InfLate}. Now we can (heuristically) see that a classical history has been reached, as the lines of real scale factor and scalar field overlap into the future imaginary $\tau$/real $\lambda$ direction.  

\begin{figure}[htbp]
\begin{minipage}{\smallWidthLeft}
\includegraphics[width=\smallWidthRight]{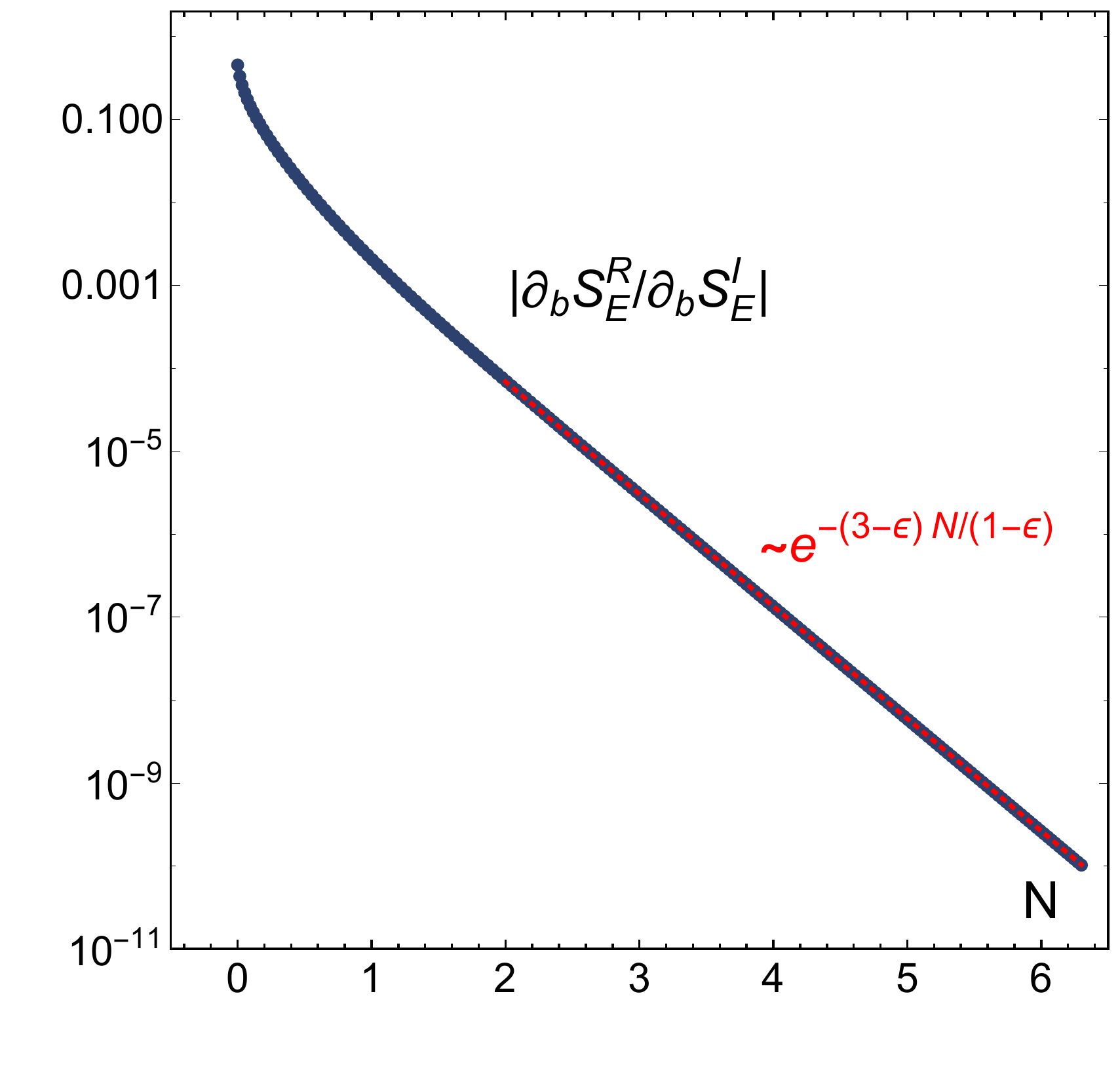}
\end{minipage}%
\begin{minipage}{\smallWidthRight}
\includegraphics[width=\smallWidthRight]{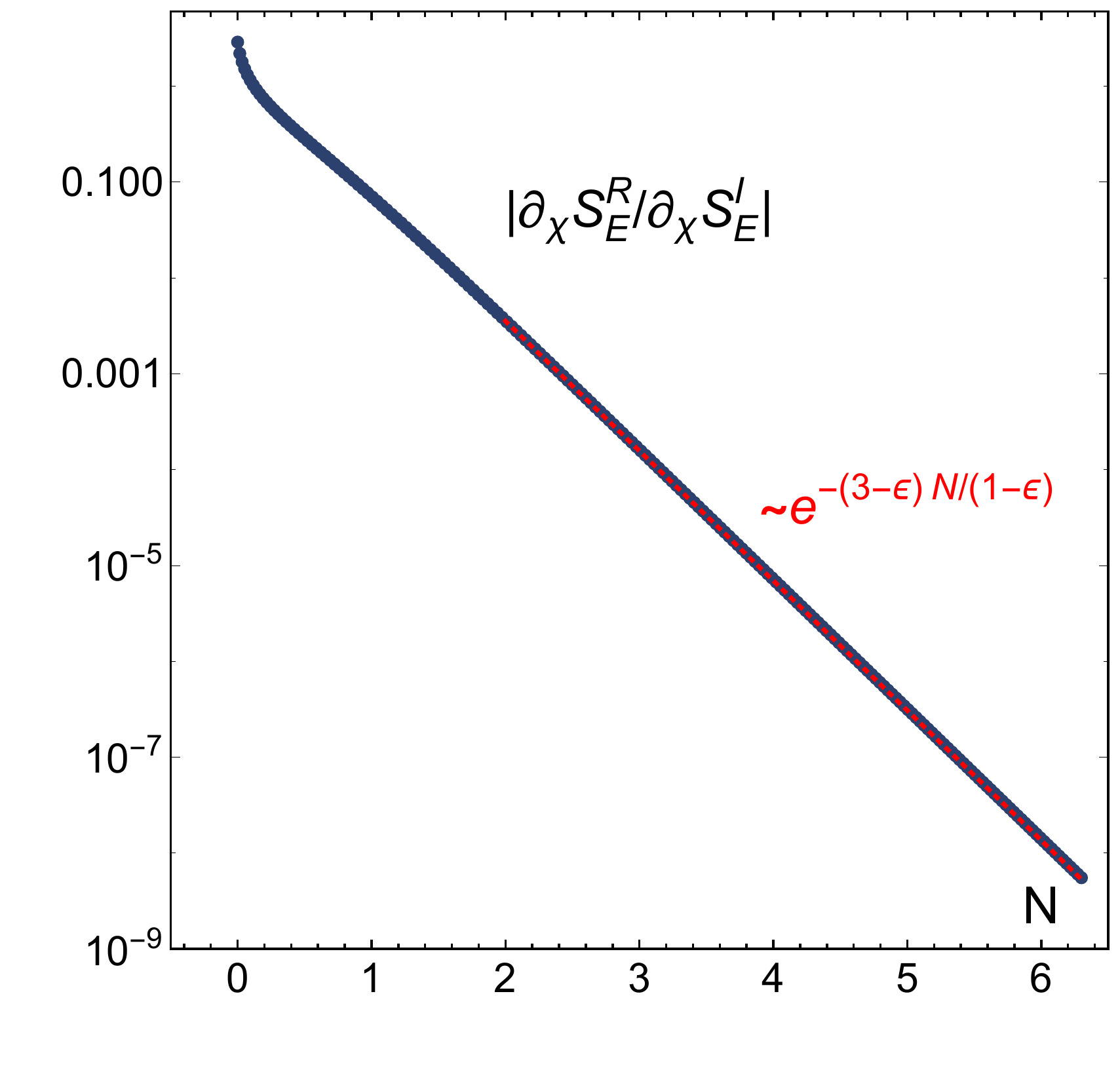}
\end{minipage}%
\caption{ \label{fig:InfWKB} For inflation with a constant slow-roll parameter $\epsilon,$ the wavefunction of the universe becomes classical exponentially fast. This is shown here by evaluating the WKB classicality conditions, which are seen to be satisfied exponentially fast in the number of e-folds $N.$ One may note how quickly the asymptotic scaling is reached.} 
\end{figure}

In order to understand the approach to classicality in more detail, we must take a look at the WKB conditions \eqref{WKBb},\eqref{WKBchi}, as discussed in section \ref{section:QC}. For this purpose, i.e. in order to estimate the $b$ and $\chi$ derivatives, we have evaluated the wavefunction for slightly shifted values of $b_{shifted} = b * (1+ 10^{-5})$ and $\chi_{shifted} = \chi + 10^{-5}$ respectively\footnote{We have checked that the finite difference estimate of the derivative is reliable by verifying that the results remain essentially unchanged when the shift is further reduced to $10^{-6}.$}. The results for the ratios $|\partial_b S_E^R/\partial_b S_E^I|$ and $|\partial_\chi S_E^R/\partial_\chi
S_E^I|$ are shown in Fig. \ref{fig:InfWKB}. Here we see something striking: not only does the wavefunction become increasingly classical as the inflationary phase proceeds, but it does so in a very precise manner, namely in proportion to the factor
\begin{equation}
\left|\frac{\partial_b S_E^R}{\partial_b S_E^I}\right|, \, \left|\frac{\partial_\chi S_E^R}{\partial_\chi S_E^I}\right|  \, \propto  \, \exp\left(-\;\frac{3-\epsilon}{1-\epsilon}\; N\right) \,.
\end{equation} 
Thus, as a function of the number of e-folds $N$, the wavefunction becomes classical exponentially fast. For small $\epsilon$ and thus approximately constant $H$, the scaling is approximately as $e^{-3N},$ i.e. the WKB conditions are satisfied in inverse proportion to the volume of space generated by inflation. Keeping in mind that a successful inflationary phase increases the volume of space by a factor of at least $e^{180}$ gives an indication as to the effectiveness of inflation in rendering spacetime classical! 

This scaling of the WKB conditions can in fact be derived analytically, in analogy with a calculation presented in the ekpyrotic context in \cite{Battarra:2014kga}. An important ingredient in that calculation are the symmetries of the action: starting from
\begin{equation}
S_E = - \int d ^4 x  \sqrt{g} \left( \frac{R}{2} - \frac{1}{2} g ^{\mu \nu} \partial _{\mu} \phi\, \partial _{\nu} \phi - V_0 e^{-c\phi} \right) \;,
\end{equation}
one can perform the following scaling and shift,
\begin{equation}
\phi  \equiv  \bar{ \phi} + \Delta \phi \;, \quad
g_{\mu\nu}  \equiv  \frac{e^{c \Delta \phi}}{ V_0} \bar{ g}_{\mu\nu} \;,\label{eq:metricscaling}
\end{equation}
which transform the action into
\begin{equation} \label{eq:actionrescaled}
S_E = -\frac{e^{c\Delta \phi}}{V_0} \int d ^4x \sqrt{ \bar{ g}} \left( \frac{ \bar{R}}{2} - \frac{1}{2} \bar{g} ^{\mu \nu} \partial _{\mu} \bar{ \phi} \partial _{\nu} \bar{ \phi} + e^{-c\bar\phi}\right) \;.
\end{equation}
Hence, if we keep $V_0=1$ so that we remain in the same theory, then the minisuperspace field equations are invariant under the transformations
\begin{eqnarray}
\bar{a} ( \bar{ \lambda}) & = & e^{c \Delta \phi / 2} \, a \left( e^{- c\, \Delta \phi /2} \bar{ \lambda} \right) \;, \label{Rescaling1}\\
\bar{ \phi}( \bar{ \lambda}) & = & \phi\left( e^{- c\, \Delta \phi /2} \bar{ \lambda} \right) + \Delta \phi \;. \label{Rescaling2}
\end{eqnarray}
The scaling/shift above takes the solution \eqref{scalingsolutionInf} into the solution
\begin{equation} \label{eq:rescalinga0}
\bar{ a} = \bar{a}_0 \, (\bar{ \lambda}) ^{1/ \epsilon} \;, \qquad \bar{a}_0 = \textrm{exp} \left( \frac{ \epsilon - 1}{ \epsilon}\frac{ c \, \Delta \phi}{2}  \right) \, a_0 \;,\qquad
V( \bar{ \phi}) =  \frac{3 - \epsilon}{ \epsilon^2} \frac{1}{ \bar{ \lambda} ^2} \;.
\end{equation}
One can see that $a_0$ is a constant of motion,
\begin{equation} \label{eq:labela0}
a_0 = a\,\left( \frac{\epsilon^2}{3-\epsilon }V \right)^{1/2 \epsilon} \;,
\end{equation}
which can be used to label the asymptotic attractor solutions of the theory. Given that the imaginary part of the Euclidean action along a classical trajectory (with $d\tau = i d\lambda$) scales as
\begin{equation}
S_E^I \sim i \, \int d \lambda\, a ^3\, V \sim i \,a_0 ^3\, (\lambda) ^{- 1 + 3/ \epsilon} \sim i \, a_0 ^3\, V ^{ \frac{1}{2} - \frac{3}{2 \epsilon}} \;,
\end{equation}
one can use the constant of motion \eqref{eq:labela0} to find
\begin{equation}
S_E^I \sim  i \, b ^3\, V( \chi) ^{1/2} \;.
\end{equation}
The scaling of the real part of the Euclidean action was found above, and it implies that 
\begin{equation}
\bar{S}_E ^{R} = e^{ c\, \Delta \phi} S_R = \left( \frac{ \bar{a}_0}{a_0} \right) ^{2 \epsilon/( \epsilon-1)} S_E ^{R} \;,
\end{equation}
so that
\begin{equation}
S_E ^{R} \sim a_0 ^{ \frac{2 \epsilon}{\epsilon - 1}} \sim b ^{\frac{2 \epsilon}{\epsilon - 1}} V( \chi) ^{1/ ( \epsilon-1)} \;.
\end{equation}
From these expressions it is now straightforward to work out the asymptotic behaviour of the WKB conditions. Noting that $\chi$ derivatives only add a constant pre factor, we have that 
\begin{equation}
\frac{\partial_\chi S_E^R}{\partial_\chi S_E^I} \propto \frac{S_E^R}{S_E^I} \sim \frac{b ^{\frac{2 \epsilon}{\epsilon - 1}} V( \chi) ^{1/ ( \epsilon-1)}}{b ^3\, V( \chi) ^{1/2}} \sim b^{\epsilon -3} \sim e^{-(\epsilon - 3)N/(\epsilon - 1)}.
\end{equation}
Meanwhile
\begin{eqnarray}
\partial _{b} S_E ^{I} &\sim& b ^2 V( \chi) ^{1/2} \sim \lambda ^{- \frac{ \epsilon - 2}{ \epsilon}} \;,\\
\partial _{b} S_E ^{R} &\sim& b ^{ \frac{ \epsilon + 1}{ \epsilon-1}} V( \chi) ^{1/( \epsilon-1)} \sim \lambda ^{- 1/ \epsilon} \;,
\end{eqnarray}
implying that 
\begin{equation}
\left| \frac{ \partial _{b} S_E ^{R}}{ \partial _{b} S_E ^{I} }\right| \sim \lambda ^{\frac{ \epsilon - 3}{ \epsilon}} \sim b^{\epsilon - 3} \sim e^{-(\epsilon - 3)N/(\epsilon - 1)} \;.
\end{equation}
This completes the analytic derivation of the asymptotic scaling of the WKB conditions.

We should point out a further consequence of the scaling and shift symmetry in \eqref{eq:metricscaling}, namely that it can be used to relate various classical histories and their corresponding instantons to another. In fact, the entire family of attractor/scaling solutions can be obtained by applying the transformations  
\begin{equation}
\chi  =  \bar{ \chi} + \Delta \chi \;, \quad
b  =  \frac{e^{c \Delta \chi/2}}{V_0^{1/2}} \bar{ b} 
\end{equation}
to the history analysed above, where the corresponding instantons then have the South Pole values $\bar\phi_{SP} + \Delta \chi.$ For all such solutions the approach to classicality will be analogous, while the relative probabilities are given by the re-scaled action \eqref{eq:actionrescaled}.

\section{Classical Spacetime from Ekpyrosis} \label{section:EkpyrInst} \label{section:ekpyrotic}

\begin{figure}[h]
\centering
\begin{minipage}{\fullWidth}
\includegraphics[width=\fullWidth]{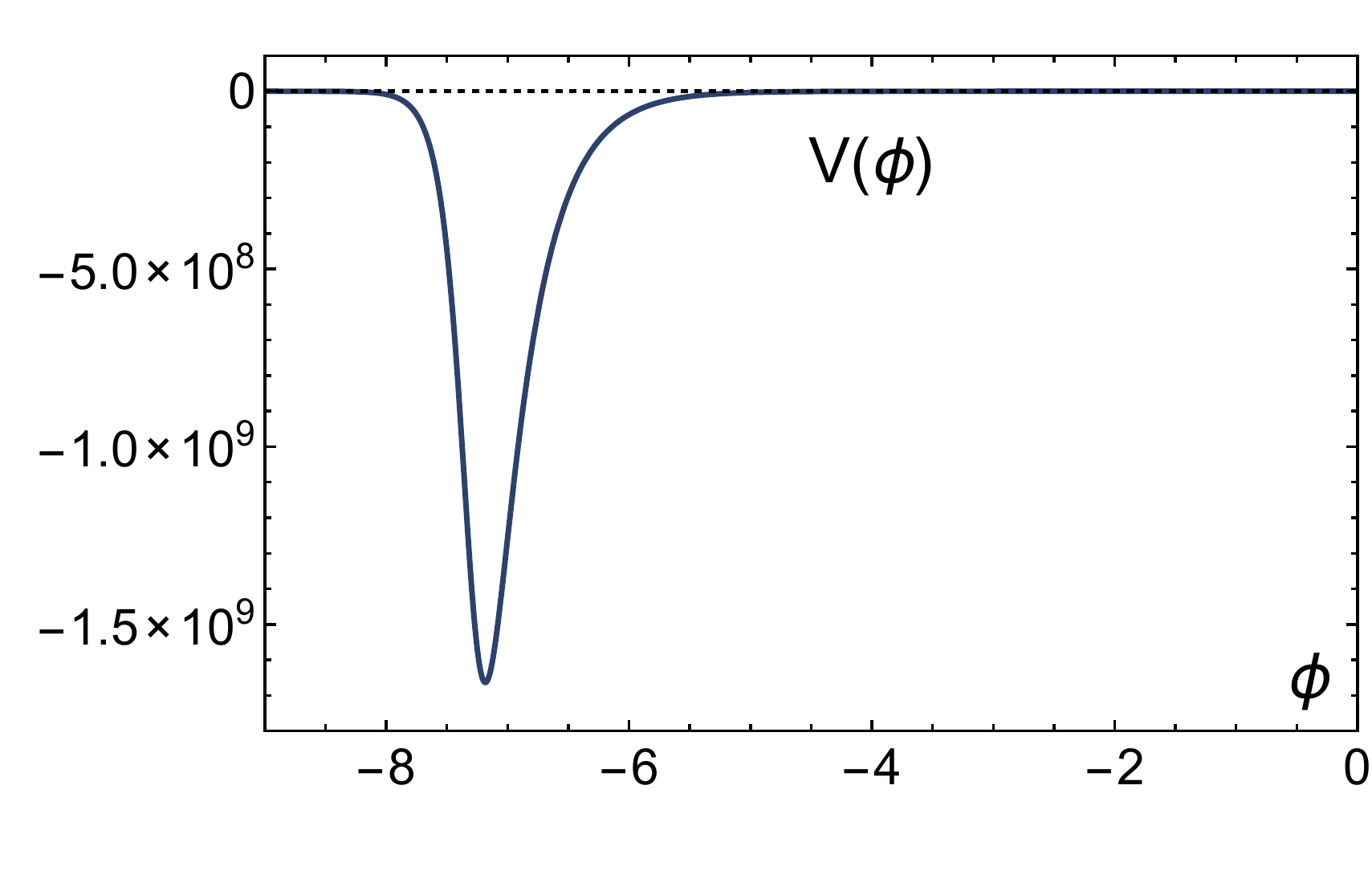}
\end{minipage}%
\caption{\label{fig:EkPot} The ekpyrotic potential. The field evolution is from right to left, with the field first rolling down the steep ekpyrotic phase while the universe is slowly contracting. As the potential turns back up towards zero, the evolution becomes dominated by the kinetic energy of the scalar field. Note that because the universe is contracting, the kinetic energy of the scalar field is blue-shifted and thus the field keeps rolling rapidly towards the left after ekpyrosis has ended.}
\end{figure}

We will now perform a similar analysis, but for a potential that is steep and negative, i.e. for an ekpyrotic phase \cite{Khoury:2001wf}. Until recently, all studies of the no-boundary proposal, and in fact of quantum cosmology in general, have been limited to positive potentials. However, it was recently discovered that with a steep and negative potential, {\it ekpyrotic instantons} exist which satisfy the no-boundary conditions and at the same time lead to a classical (contracting) universe \cite{Battarra:2014xoa,Battarra:2014kga}. For the remainder of this paper, we will study this new type of solutions in detail, especially regarding the approach of classicality. It may at first sound surprising that no-boundary instantons can be found leading to a contracting universe. The South Pole region of no-boundary instantons is typically interpreted as the region where space and time tunnel out of ``nothing'', suggesting that afterwards the universe should first undergo a period of expansion during which more space gets created. Ekpyrotic instantons get around this apparent paradox due to the fact that the South Pole region is a (large) region of Euclidean flat space which then smoothly interpolates (via a region where the fields are fully complex) to an increasingly classical contracting Lorentzian universe. An interesting feature of these solutions is that they have a very high relative probability, much higher than that of realistic inflationary universes. Thus, if the potential contains both regions that are positive and flat, and regions that are negative and steep, the wavefunction of the universe will be dominated by contracting universes. This by itself is already a good motivation to study their properties in more detail. Of course, such universes are only realistic if they can revert from contraction to expansion eventually, i.e. they must undergo a bounce. We will leave this important issue for future work. In the present paper, we will analyse how such contacting ekpyrotic universe become classical, and whether or not they preserve their classicality when the ekpyrotic phase comes to an end. Moreover, we will look at the implications of adding a dark energy plateau, as envisaged in cyclic models of the universe \cite{Steinhardt:2001st}.

\subsection{The ekpyrotic and kinetic phases} \label{section:ekkin}

We will consider a potential of the form
\begin{equation}
V(\phi) = - \frac{V_0}{ e^{c_1 \phi} + e^{-c_2 (\phi + c_3)}},
\end{equation}
where $V_0> 0$ and $c_{1,2,3}$ are constants. For our numerical example we have chosen $V_0=1, c_1 = 3, c_2 = 8, c_3 = 10$ - see Fig. \ref{fig:EkPot}. For $\phi \gtrsim -7$ the potential is well-approximated by the pure ekpyrotic potential $V=- e^{-3\phi},$ but we have added an additional term which effectively switches the potential off at large negative $\phi$ values, thus allowing the ekpyrotic phase to come to an end. This will allow us to analyse what happens when the universe becomes dominated by the kinetic energy of the scalar field \cite{Lehners:2008vx}. 

\begin{figure}[htbp]
\begin{minipage}{\smallWidthLeft}
\includegraphics[width=\smallWidthRight]{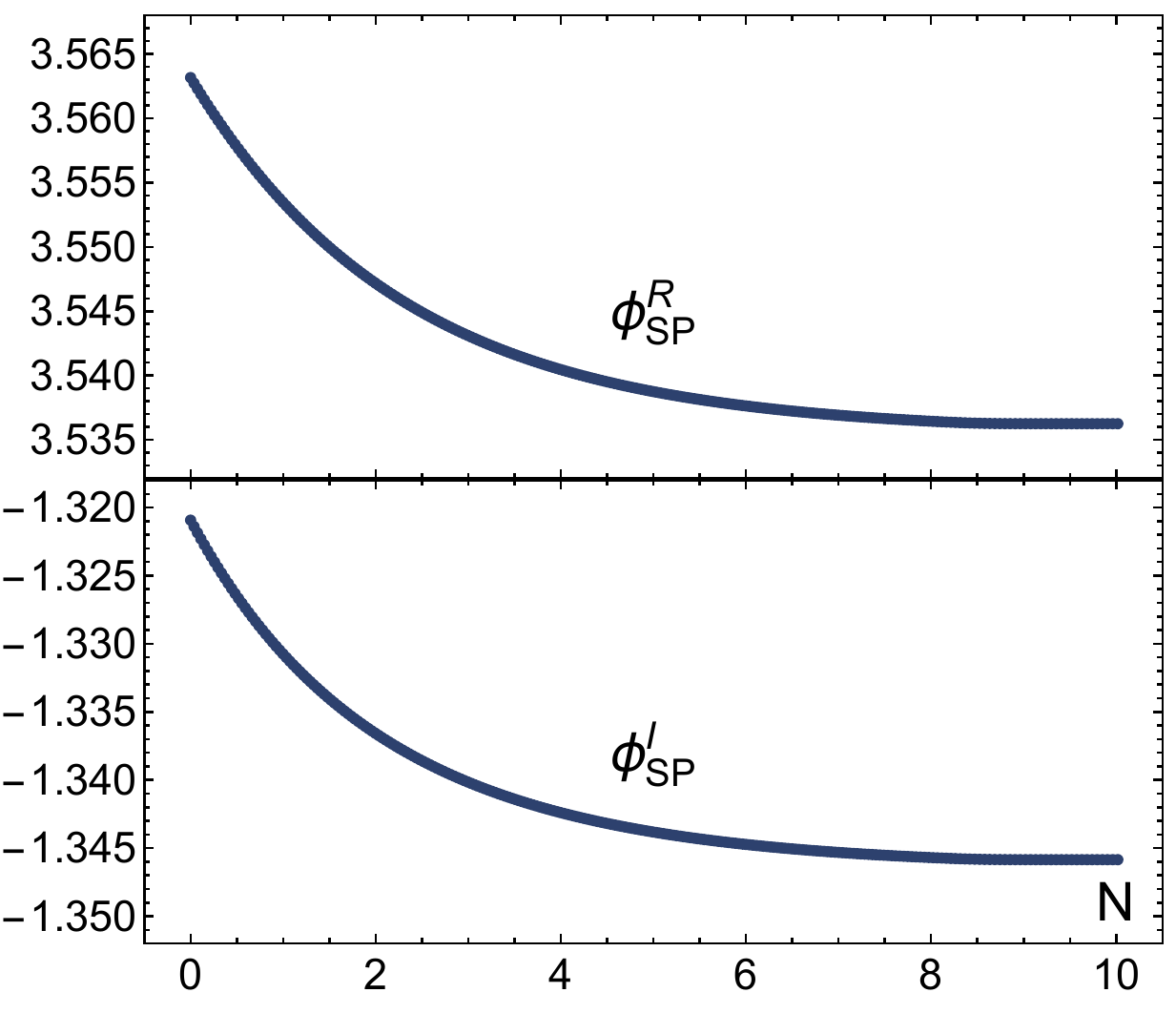}
\end{minipage}%
\begin{minipage}{\smallWidthRight}
\includegraphics[width=\smallWidthRight]{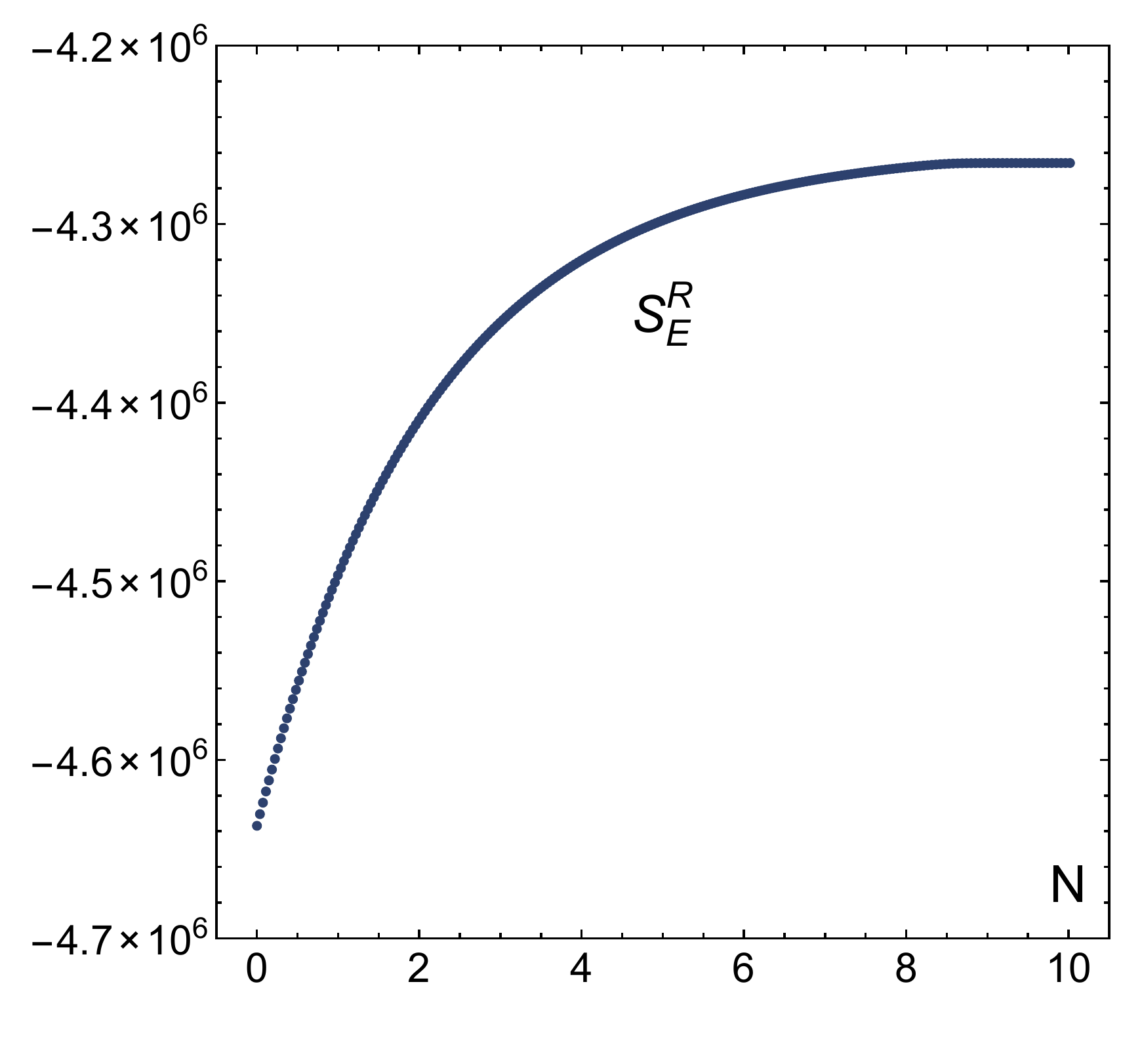}
\end{minipage}%
\caption{ \label{fig:EkPhi0ReS} Left panel: the South Pole values $\phi_{SP}$ of the scalar field for the instantons corresponding to the classical solution specified by Eqs. \eqref{EkIC1} - \eqref{EkIC2}. Right panel: the corresponding real part of the Euclidean action. One can see that as the ekpyrotic phase proceeds, the values of $\phi_{SP}$ and $S_E^R$ stabilise, which is a sign that the wavefunction has become WKB --``classical''.} 
\end{figure}

For our numerical computations, we have chosen a classical solution with initial conditions
\begin{eqnarray}
c & = 3, \qquad \epsilon & = \frac{9}{2} \\ a(\lambda_i)  & = 100, \qquad \dot{a}(\lambda_i) & = - \left( -1 + \frac{a(\lambda_i)^2}{3}(\frac{1}{2} \dot\phi(\lambda_i)^2 + V(\lambda_i))\right)^{1/2} \label{EkIC1} \\
\phi(\lambda_i) = & 0, \qquad \dot{\phi}(\lambda_i) & = - \left( \frac{2\epsilon V(\lambda_i)}{3-\epsilon} \right)^{1/2}\, . \label{EkIC2}
\end{eqnarray}
The evolution starts in the ekpyrotic phase, where the fast-roll parameter
\begin{equation}
\epsilon = \frac{V_{,\phi}^2}{2 V^2} = \frac{c^2}{2}
\end{equation}
is constant. The evolution quickly reaches the asymptotic scaling solution
\begin{equation}
a=a_0 (-\lambda)^{1/\epsilon}, \qquad \phi = \sqrt{\frac{2}{\epsilon}} \ln \left( - \sqrt{\frac{\epsilon^2 V_0}{3-\epsilon}} \, \lambda\right), \qquad V = - \frac{\epsilon - 3}{\epsilon^2\lambda^2}\,. \label{scalingsolution}
\end{equation} 
When $\epsilon > 3$ this solution corresponds to an ekpyrotic attractor solution. For $\phi \lesssim -7$ the potential becomes unimportant, and the ekpyrotic phase goes over into the kinetic phase, during which the solution is given by
\begin{equation}
a=a_c (\lambda_c-\lambda)^{1/3}, \qquad \phi = \sqrt{\frac{2}{3}} \ln \left( \lambda_c-\lambda \right), \qquad V \approx 0\,, \label{kineticsolution}
\end{equation} 
where $\lambda_c$ corresponds to the time when the universe crunches.

\begin{figure}[htbp]
\begin{minipage}{\smallWidthLeft}
\includegraphics[width=\smallWidthRight]{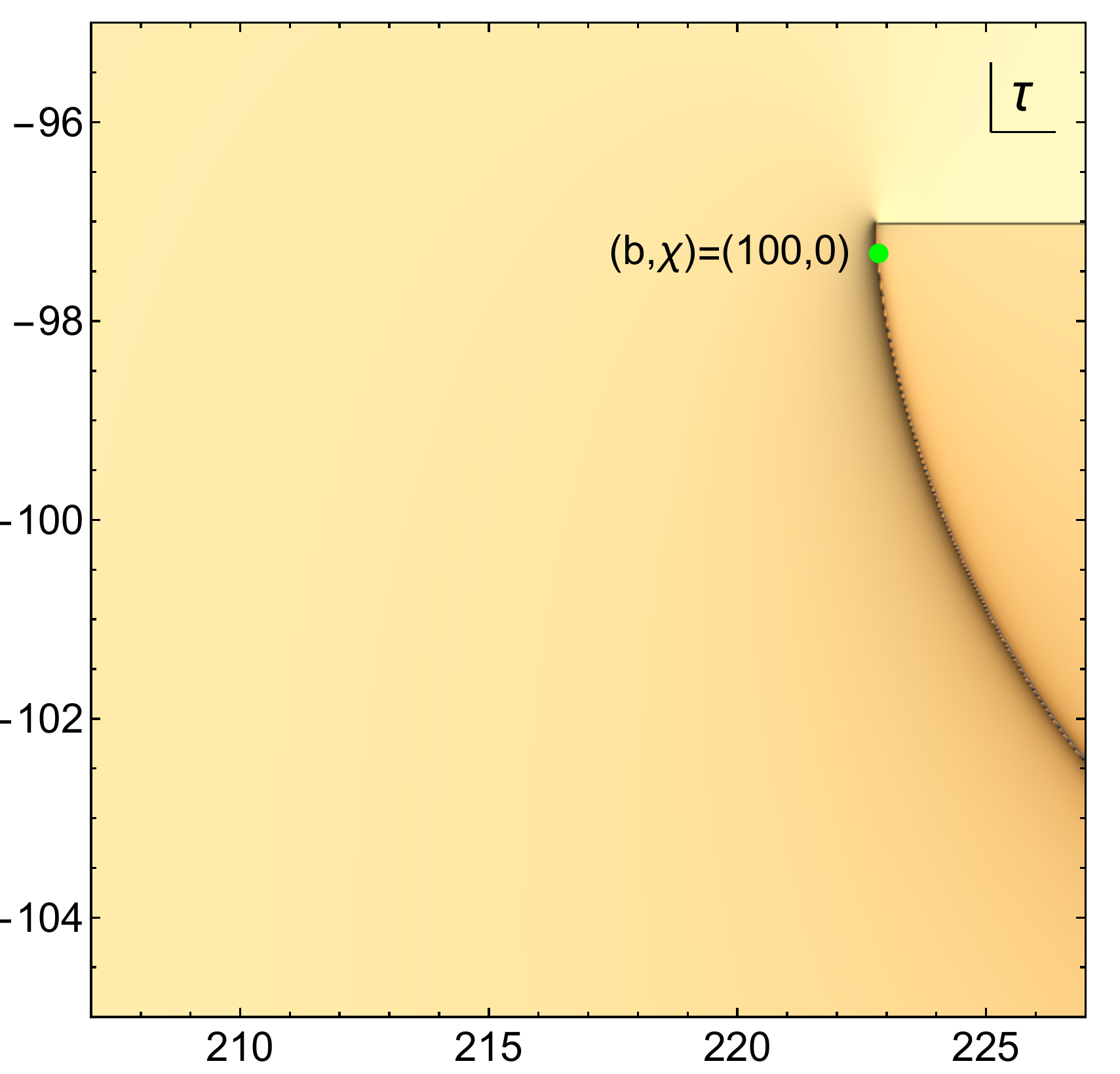}
\end{minipage}%
\begin{minipage}{\smallWidthRight}
\includegraphics[width=\smallWidthRight]{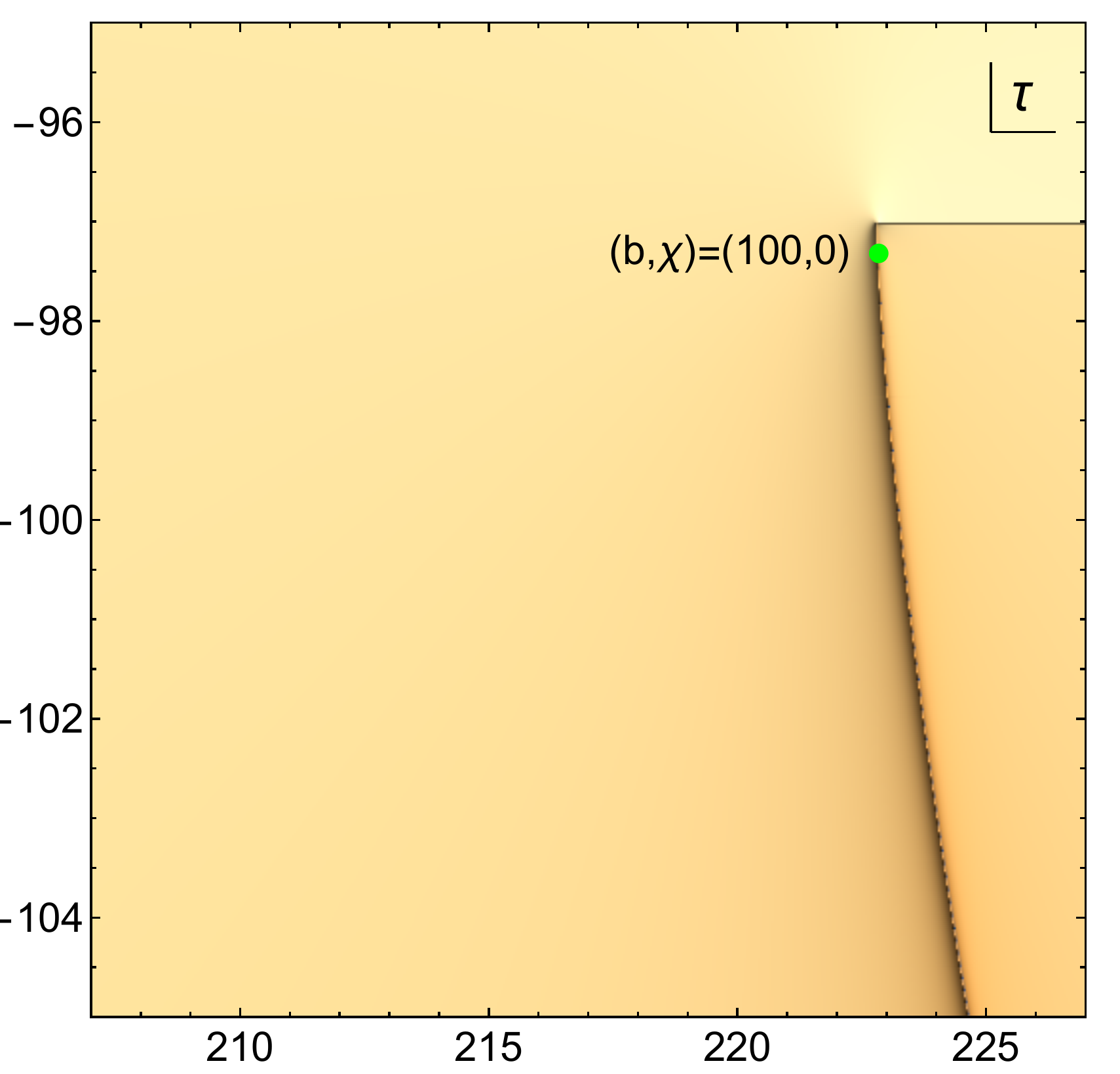}
\end{minipage}%
\caption{ \label{fig:EkEarly} A visual representation of the instanton corresponding to the values $(b=100,\chi=0)$. This instanton is characterised by the value $\phi_{SP}=3.563-1.321 i.$ This figure was evaluated in the same manner as Fig. \ref{fig:InfEarly}.} 
\end{figure}

\begin{figure}[htbp]
\begin{minipage}{\smallWidthLeft}
\includegraphics[width=\smallWidthRight]{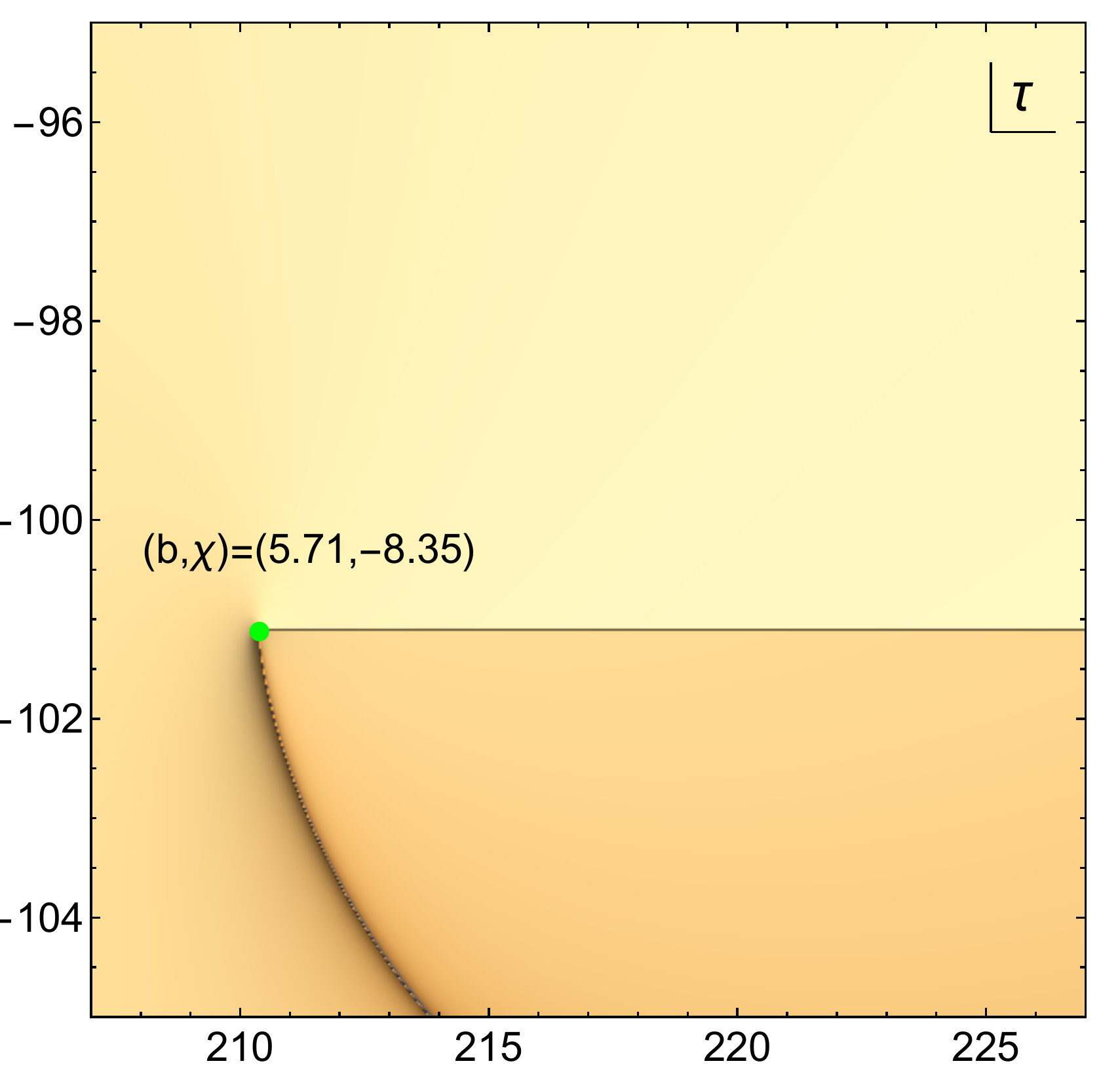}
\end{minipage}%
\begin{minipage}{\smallWidthRight}
\includegraphics[width=\smallWidthRight]{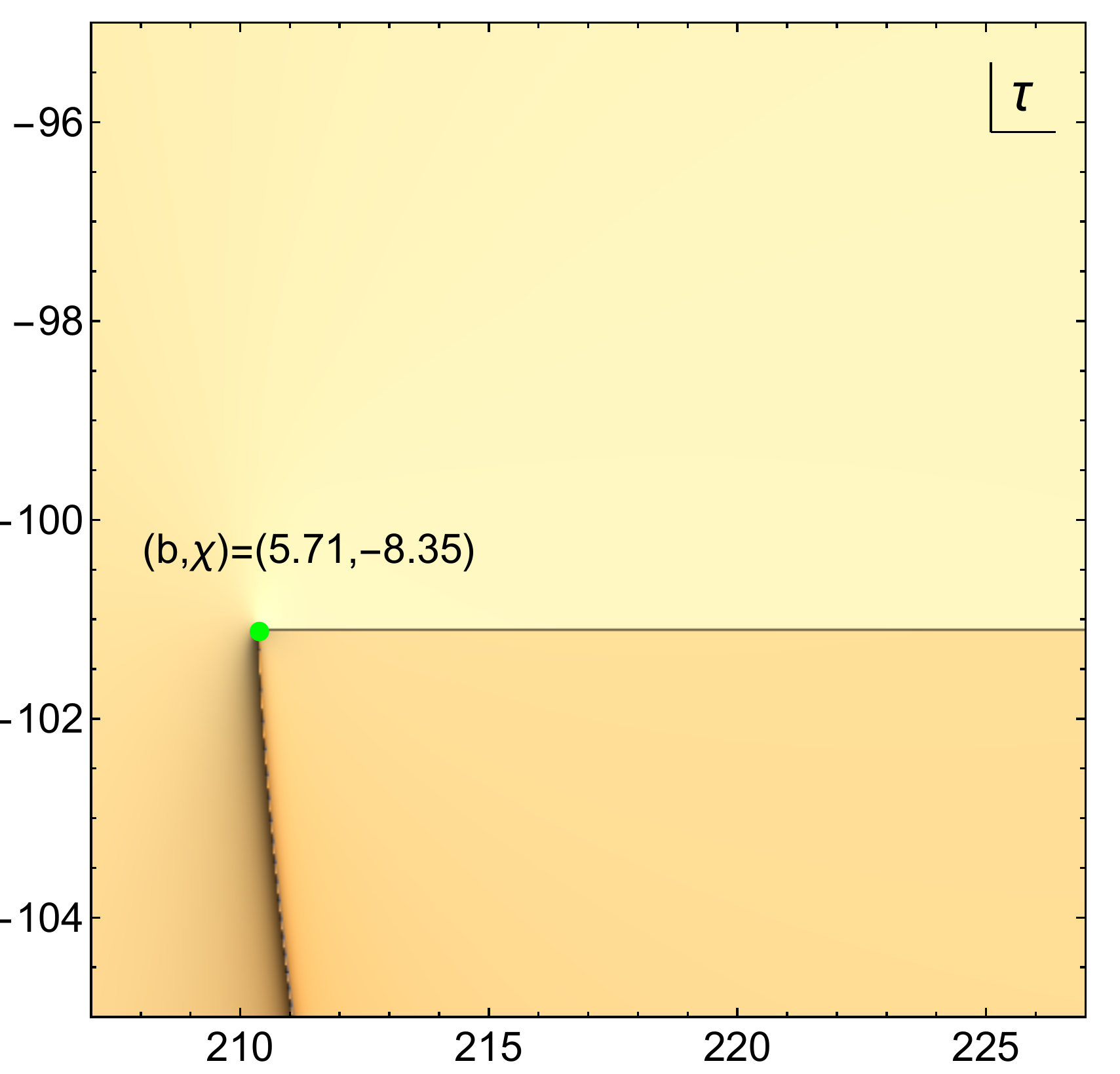}
\end{minipage}%
\caption{ \label{fig:EkLate} A visual representation of the instanton corresponding to the values $(b=5.71,\chi=-8.35)$. This instanton is characterised by the value $\phi_{SP}=3.536-1.346 i.$ Note that the lines of real scale factor and real scalar field become aligned and vertical just before the crunch. The range of $\lambda$ during the ekpyrotic and kinetic phases is very small compared to the range of the scalar field.} 
\end{figure}

Then, in order to evaluate the properties of the no-boundary wavefunction, we have evaluated the wavefunction for $250$ values of $(b,\chi)$ ranging between $(100, 0)$ and $(5.71, -8.35)$ along the classical solution, corresponding to a range of about 10 e-folds of contraction. The corresponding South Pole values of the scalar field, and real parts of the Euclidean action, are shown in Fig. \ref{fig:EkPhi0ReS}. As can be seen, these values stabilise after a few e-folds, thus indicating that a classical history is reached. Before analysing the approach to classicality in more detail, it is instructive to take a look at the actual instantons, in order to contrast them with the inflationary case discussed in the previous section. The first and last instanton in the series are shown in Figs. \ref{fig:EkEarly} and \ref{fig:EkLate} respectively, in analogy with Figs. \ref{fig:InfEarly} and \ref{fig:InfLate}. There are a number of features to note: the first is that the first and last instanton in the series don't differ substantially in their shape. However, note that the positions of the lines of real $a$ and $\phi$ shift by a significant amount. Second, the lines of real scale factor and scalar field come to an end. This is the moment where the crunch occurs ($\lambda=\lambda_c$), and this is also the moment beyond which one cannot continue the analysis in the theory considered in the present paper. Third, the ekpyrotic and kinetic phases occur over a small time period (i.e. a small range of $\lambda$), even though the scalar field range is quite large - this feature makes the visualisation a little less intuitive, as most of the evolution is condensed to being very near the crunch in the figures. Finally, note that the lines of real scale factor and scalar field become increasingly aligned and vertical as the ekpyrotic phase proceeds.  

\begin{figure}[htbp]
\begin{minipage}{\smallWidthLeft}
\includegraphics[width=\smallWidthRight]{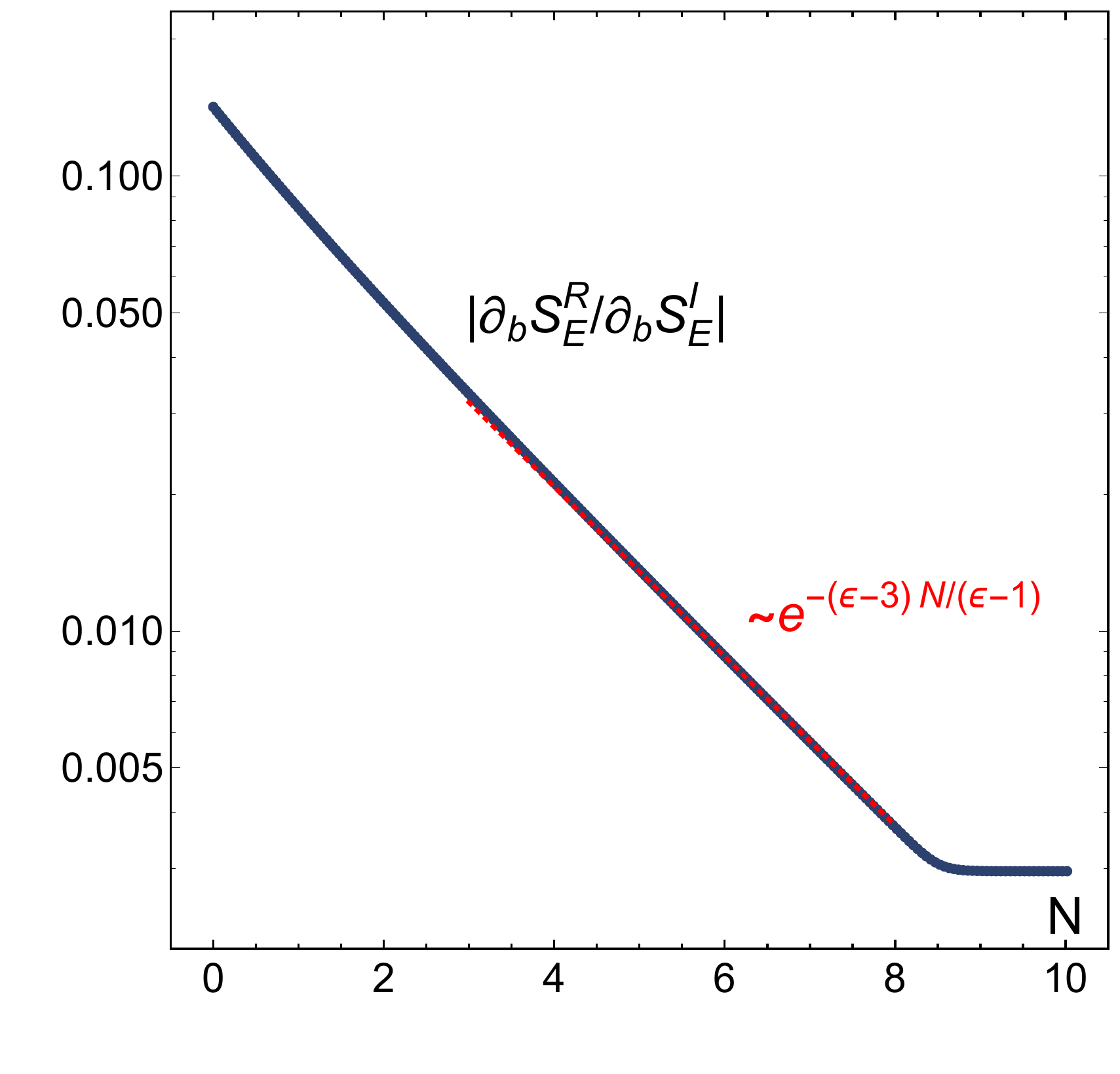}
\end{minipage}%
\begin{minipage}{\smallWidthRight}
\includegraphics[width=\smallWidthRight]{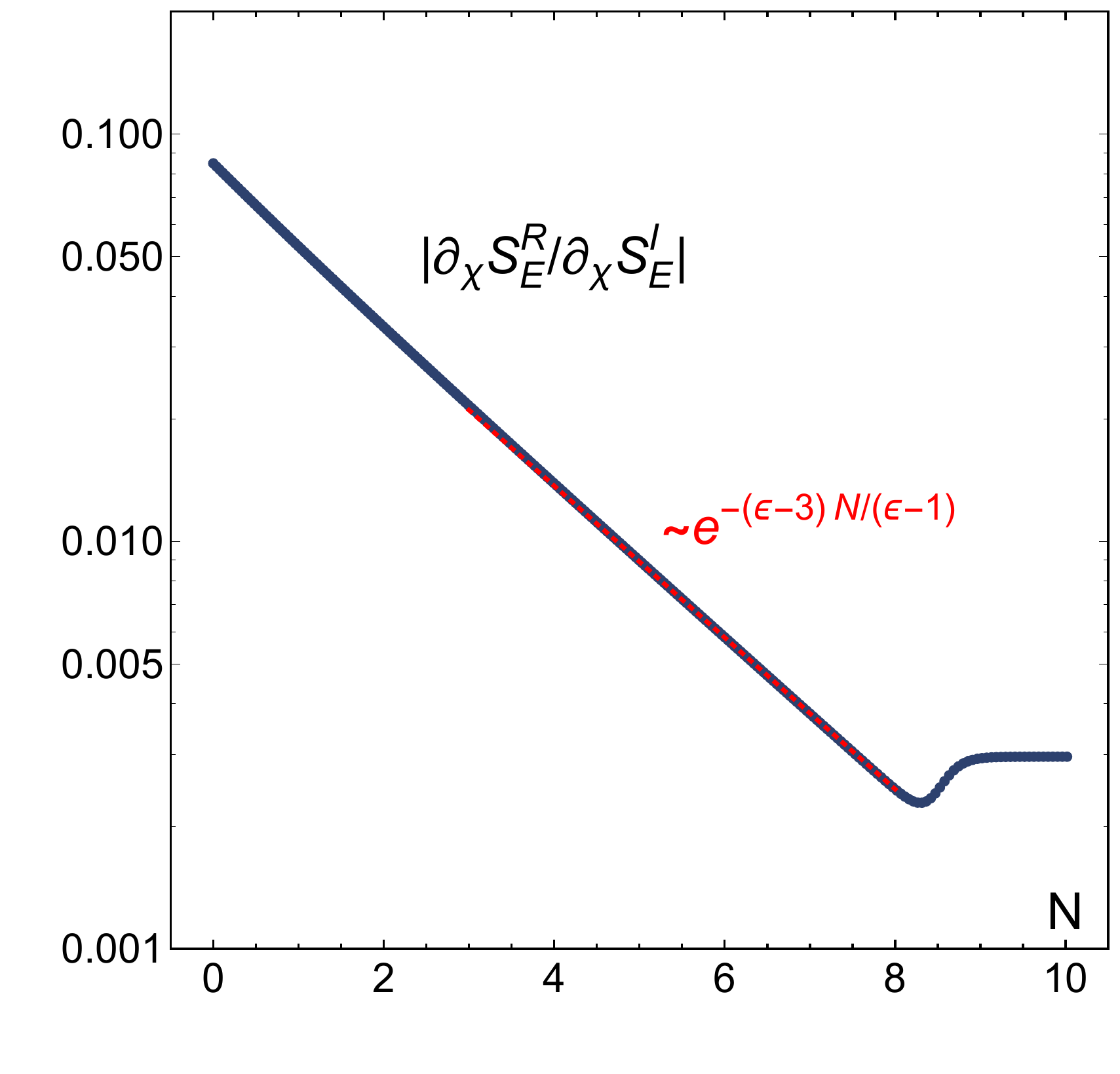}
\end{minipage}%
\caption{ \label{fig:EkWKB} The WKB classicality conditions for an ekpyrotic phase followed by a kinetic phase. The universe becomes classical exponentially fast in the number of e-folds of ekpyrotic contraction, while the level of classicality that is reached in this way is essentially preserved during the subsequent kinetic phase.} 
\end{figure}

Fig. \ref{fig:EkWKB} presents plots of the WKB classicality conditions along the contracting classical history. These plots confirm that classicality is reached: during the ekpyrotic phase, in complete analogy with the inflationary case, the WKB conditions are satisfied exponentially fast in the number of e-folds $N,$
\begin{equation}
\left|\frac{\partial_b S_E^R}{\partial_b S_E^I}\right|, \, \left|\frac{\partial_\chi S_E^R}{\partial_\chi S_E^I}\right|  \, \propto  \, \exp\left(-\;\frac{3-\epsilon}{1-\epsilon}\; N\right) \,.
\end{equation} 
Again, this scaling is attained very quickly. Moreover, the asymptotic exponential scaling with $N$ can be derived analytically, as first shown in \cite{Battarra:2014kga} (we will not repeat the derivation here, which is analogous to the one presented for the inflationary case in section \ref{section:inflation} in any case). If the equation of state during ekpyrosis is ultra-stiff, i.e. if $\epsilon \gg 3,$ classicality is reached as $\sim e^{-N}.$ Thus, remembering that for inflation the scaling was found to be $\sim e^{-3N},$ one can see that inflation is even more efficient in rendering spacetime classical, but the important point is that in both cases the scaling is {\it exponential} with $N.$

The plot in Fig. \ref{fig:EkWKB} shows another important property: as the ekpyrotic phase goes over into the kinetic phase at $N \approx 8.5,$ the WKB conditions reach a constant value. In other words, the level of classicality reached during the ekpyrotic phase is preserved during the kinetic phase, in the approach to the crunch. One can also understand this scaling analytically: during the kinetic phase the potential $V$ becomes unimportant, so that the on-shell action is (asymptotically) given by   
\begin{equation}
S_E ^{on-shell} =  -12 \pi ^2 \int d \tau\,  a \approx -12 \pi \int d \tau \, a_0 \, (\lambda_c - \lambda)^{1/3}  \;,
\end{equation}
where $a_0$ will contain a small imaginary part. Thus the real and imaginary parts of the Euclidean action are proportional to one another, and hence the WKB conditions $\partial S_E^R/\partial S_E^I$ become constant.

In order to obtain a realistic cosmology, two additional ingredients are required: the first is that a mechanism must be added which allows for the generation of nearly scale-invariant and nearly Gaussian curvature perturbations. A number of such mechanisms exist, all involving the addition of a second scalar field - see e.g. \cite{Li:2013hga,Qiu:2013eoa,Fertig:2013kwa,Ijjas:2014fja}. The second missing ingredient is a description of the bounce into an expanding phase. This crucial aspect of bouncing models remains incompletely understood at present, but several promising ideas exist for incorporating a bounce. One possibility is that the bounce is classically singular, and must be described in quantum gravity \cite{Khoury:2001wf,Ashtekar:2006wn,Bars:2011aa}. A second possibility is that the bounce may be classically non-singular and describable in an effective classical theory \cite{Buchbinder:2007ad,Creminelli:2007aq,Easson:2011zy,Cai:2012va,Koehn:2013upa,Battarra:2014tga,Battefeld:2014uga,Qiu:2015nha}. In that case, it is certainly imperative that spacetime become highly classical in the approach to the bounce. Here we have demonstrated that this is exactly what happens.

\subsection{The cyclic potential}

\begin{figure}[htbp]
\begin{minipage}{\smallWidthLeft}
\includegraphics[width=\smallWidthRight]{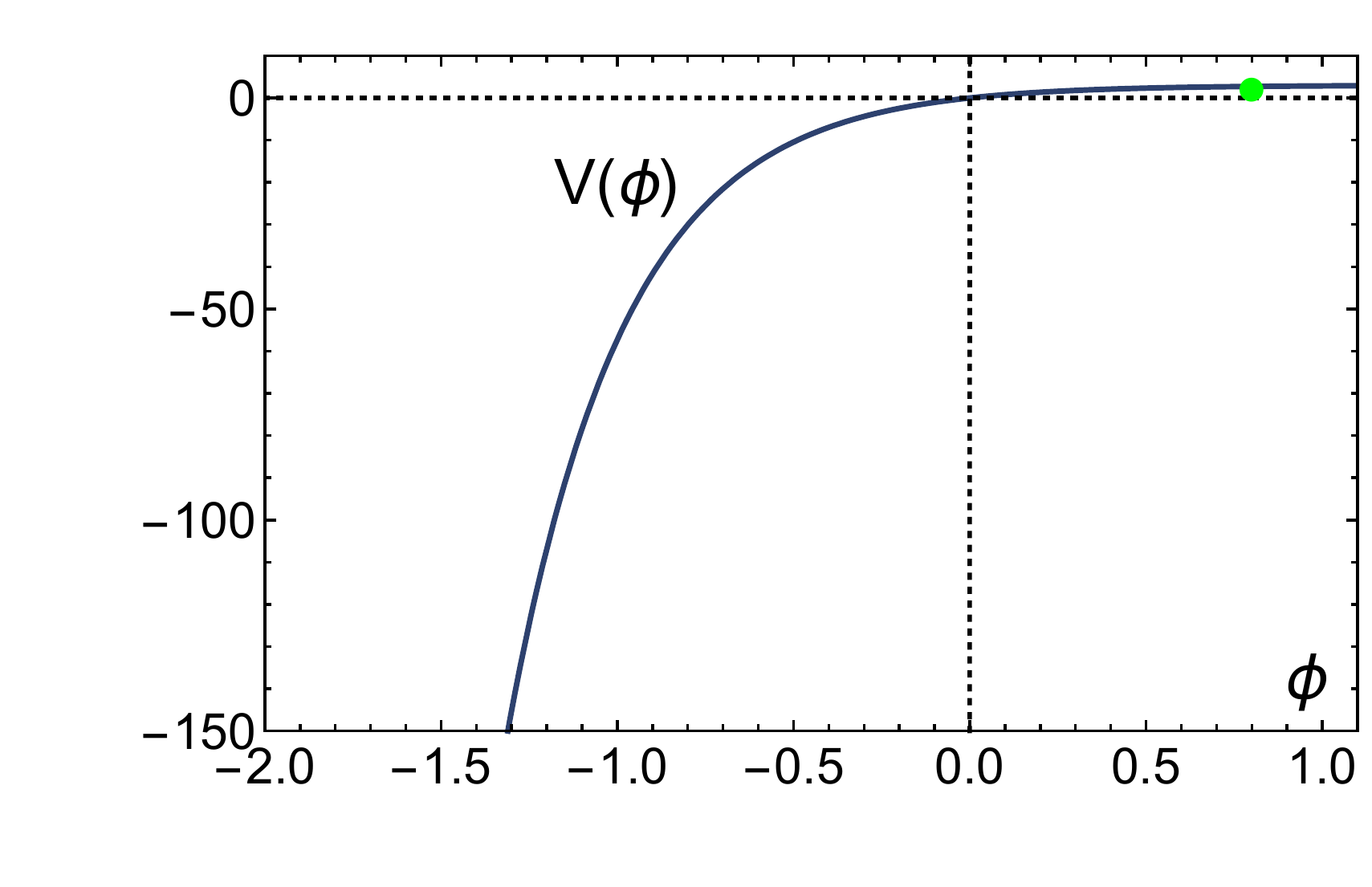}
\end{minipage}%
\begin{minipage}{\smallWidthRight}
\includegraphics[width=\smallWidthRight]{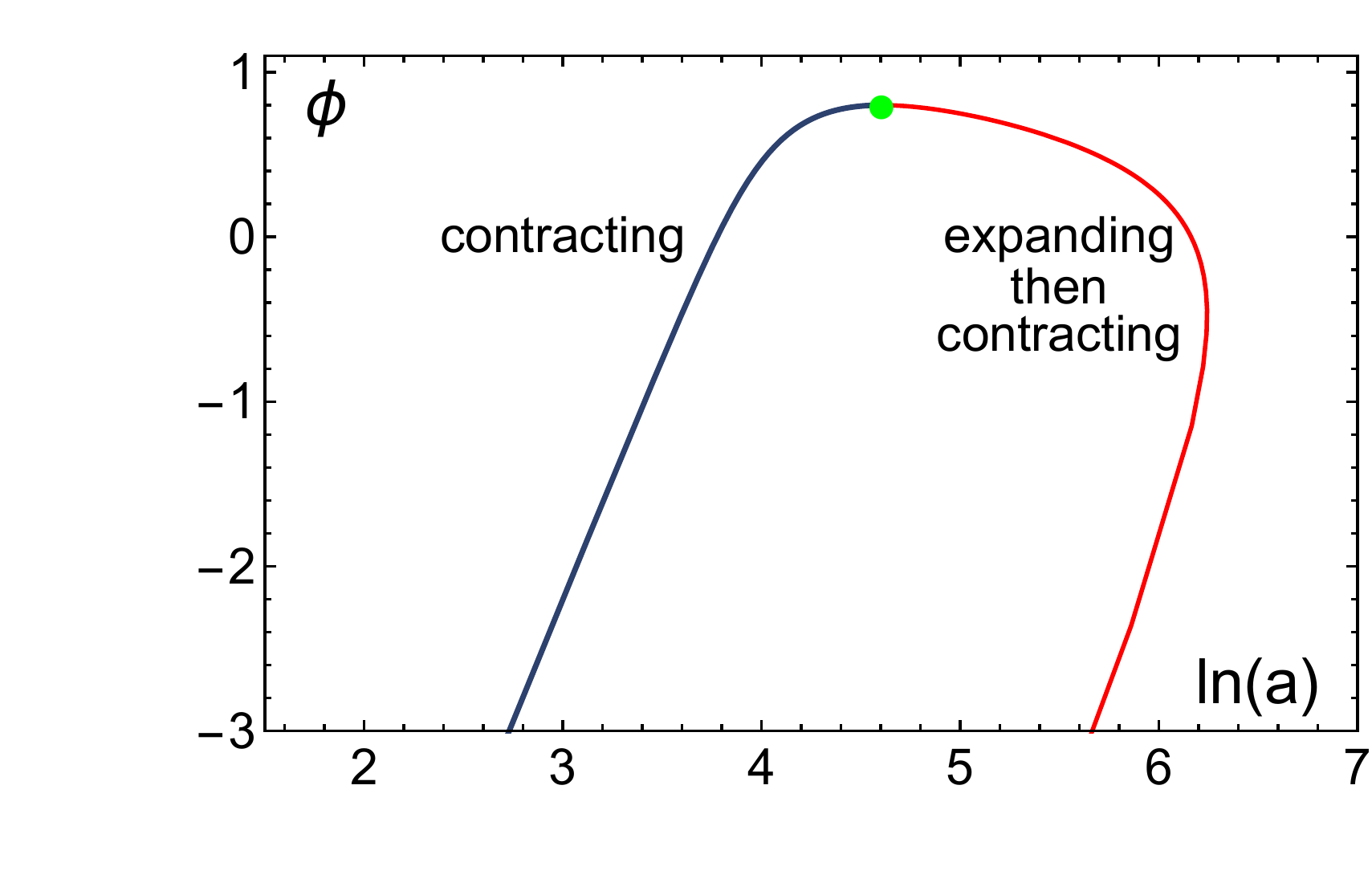}
\end{minipage}%
\caption{ \label{fig:CycPot} Left panel: the potential for a cyclic universe contains a dark energy plateau. Right panel: we will be interested in two different classical solutions, both starting from the green dot on the potential. One history is always contracting, while the second one is initially expanding, and then later on reverts to contraction as the potential becomes negative. Note that these two histories are not the time-reverse of each other.} 
\end{figure}

The cyclic universe is a framework for a more complete cosmological model, including alternating contracting and expanding phases, and thus linking the early and the late universe with each other \cite{Steinhardt:2001st,Lehners:2008vx}. The central idea in going from an ekpyrotic model to a cyclic one is that in the future, the current dark energy period (modelled as quintessence) will come to an end when the scalar potential decreases and becomes negative. This will cause the universe to stop expanding and to enter a new ekpyrotic contraction period. In this way a cyclic behaviour is achieved, with each cycle setting up the initial conditions for the next one\footnote{If the perturbations are generated by having an unstable potential, as envisaged in \cite{Finelli:2002we,Lehners:2007ac,Koyama:2007mg,Lehners:2007wc}, then the issue of initial conditions is more involved, as described in \cite{Lehners:2008qe,Lehners:2009eg,Lehners:2011ig}. Here we will restrict to models where the potential is stable, i.e. we implicitly assume the perturbation generating mechanism described in \cite{Li:2013hga,Qiu:2013eoa,Fertig:2013kwa,Ijjas:2014fja}.}. In that sense, a cyclic universe improves the issue of initial conditions. However, what is left unexplained is how the initial conditions for the first cycle were set up, and how space and time came to behave classically (these questions are relevant even if there were an infinite number of cycles). In order to answer these questions, a theory of initial conditions is needed, and in this paper we will analyse the issue of classicality from the point of view of the no-boundary proposal.

\begin{figure}[htbp]
\begin{minipage}{\smallWidthLeft}
\includegraphics[width=\smallWidthRight]{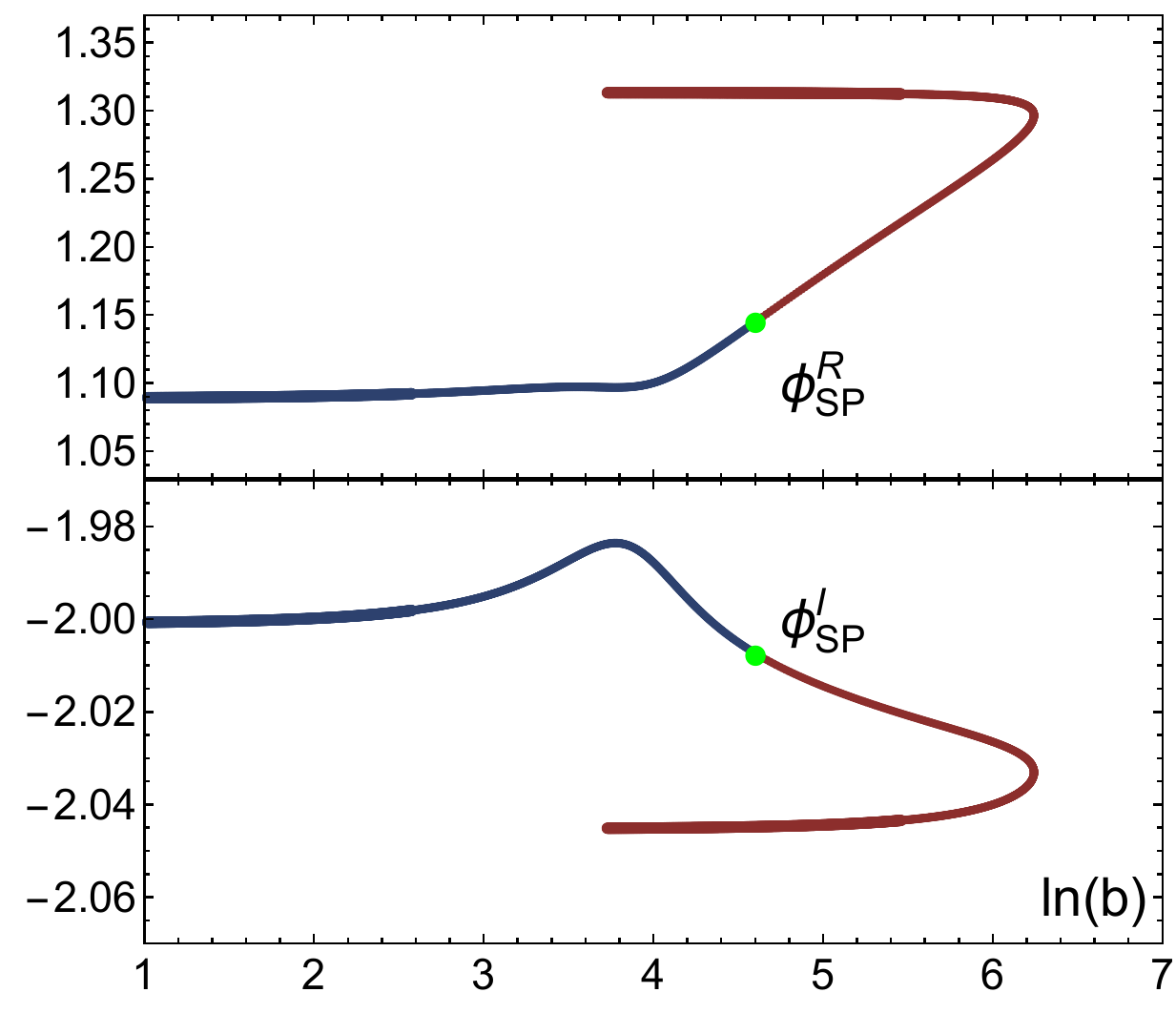}
\end{minipage}%
\begin{minipage}{\smallWidthRight}
\includegraphics[width=\smallWidthRight]{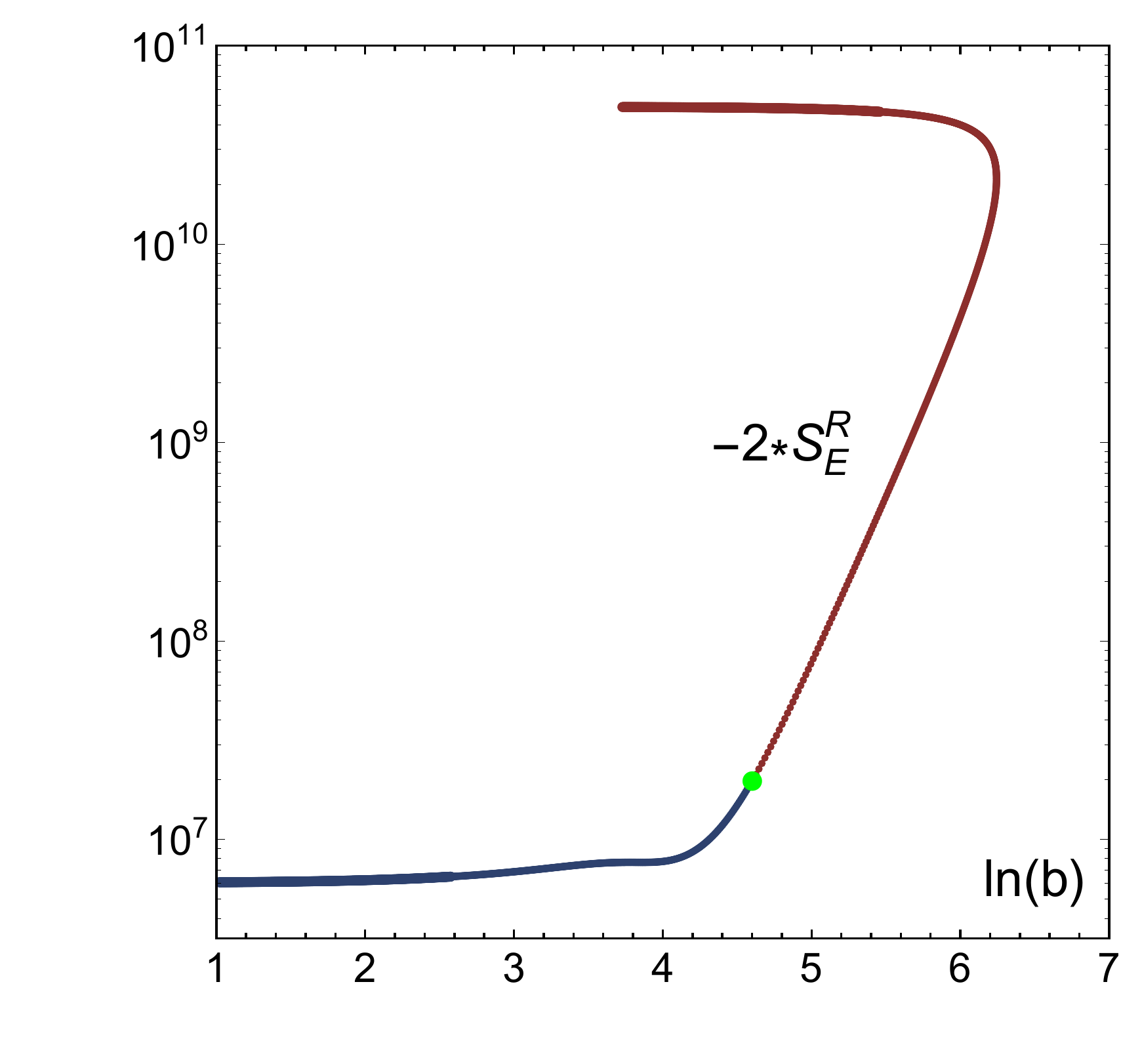}
\end{minipage}%
\caption{ \label{fig:CycPhi0ReS} Left panel: the South Pole values of the instantons associated with the classical histories shown in Fig. \ref{fig:CycPot}. Right panel: the logarithm of the relative probability (given by minus two times the Euclidean action) for the same instantons. During the contracting part of these histories, both the values of $\phi_{SP}$ and $-2S_E^R$ stabilise, indicating that the wavefunction has become WKB classical.} 
\end{figure}

The potential for the cyclic universe is shown in the left panel of Fig. \ref{fig:CycPot}. It was established in \cite{Battarra:2014kga} that on the dark energy plateau, two types of instantons exist: inflationary ones, for which the universe becomes classical due to the low-energy inflationary expansion, and ekpyrotic ones, for which the universe becomes classical due to the subsequent rolling down the steep negative part of the potential. Moreover, it was shown that the latter instantons have a higher relative probability. For this reason (and also because inflationary instantons were already discussed above) we will focus on the latter ones. We will illustrate the implications of the no-boundary proposal by contrasting two possible classical histories, which however start out at the same values of the scale factor and scalar field, but where one solution is contracting, while the other one is initially expanding and only later starts contracting, as envisaged in the cyclic scenario - see the right panel in Fig. \ref{fig:CycPot}. Note that the first type of solution would however also lead to a cyclic evolution later on, so that both are valid histories, and one may ask which one is preferred, and whether they have different properties. More specifically, the initial conditions we are interested in are given by
\begin{eqnarray}
V(\phi) & = 3 \, \left( 1 - e^{- 3 \phi} \right), \\ a(\lambda_i)  & = 100, \qquad \dot{a}(\lambda_i) &  = \pm \left( -1 + \frac{a(\lambda_i)^2}{3}(\frac{1}{2} \dot\phi(\lambda_i)^2 + V(\lambda_i))\right)^{1/2} \label{CycIC1} \\
\phi(\lambda_i) = & \frac{4}{5}, \qquad \dot{\phi}(\lambda_i) & = - 10^{-6}\, . \label{CycIC2}
\end{eqnarray}
These two solutions are not the time reverse of each other, as the initial value of $H$ is flipped, but not that of $\dot\phi.$ We have evaluated the no-boundary wavefunction with both classical histories as arguments -- more specifically, we have calculated the relevant instantons for $1500$ values between $(b=100, \chi=4/5)$ and $(b=2.766, \chi=-8.110)$ for the contracting history, and between $(b=100, \chi=4/5)$ and $(b=41.60, \chi=-3.686)$ for the initially expanding history. The corresponding South Pole values of the scalar field, as well as (the logarithm of) the relative probabilities are shown in Fig. \ref{fig:CycPhi0ReS}. These figures suggest that during the contracting part of the evolution, a classical history is reached. Moreover, the right panel shows that the initially expanding history (in red) is significantly likelier (by an astonishing factor of almost $e^{10^{11}}$) than the contracting one, according to the probability measure associated with the no-boundary proposal.

\begin{figure}[htbp]
\begin{minipage}{\smallWidthLeft}
\includegraphics[width=\smallWidthRight]{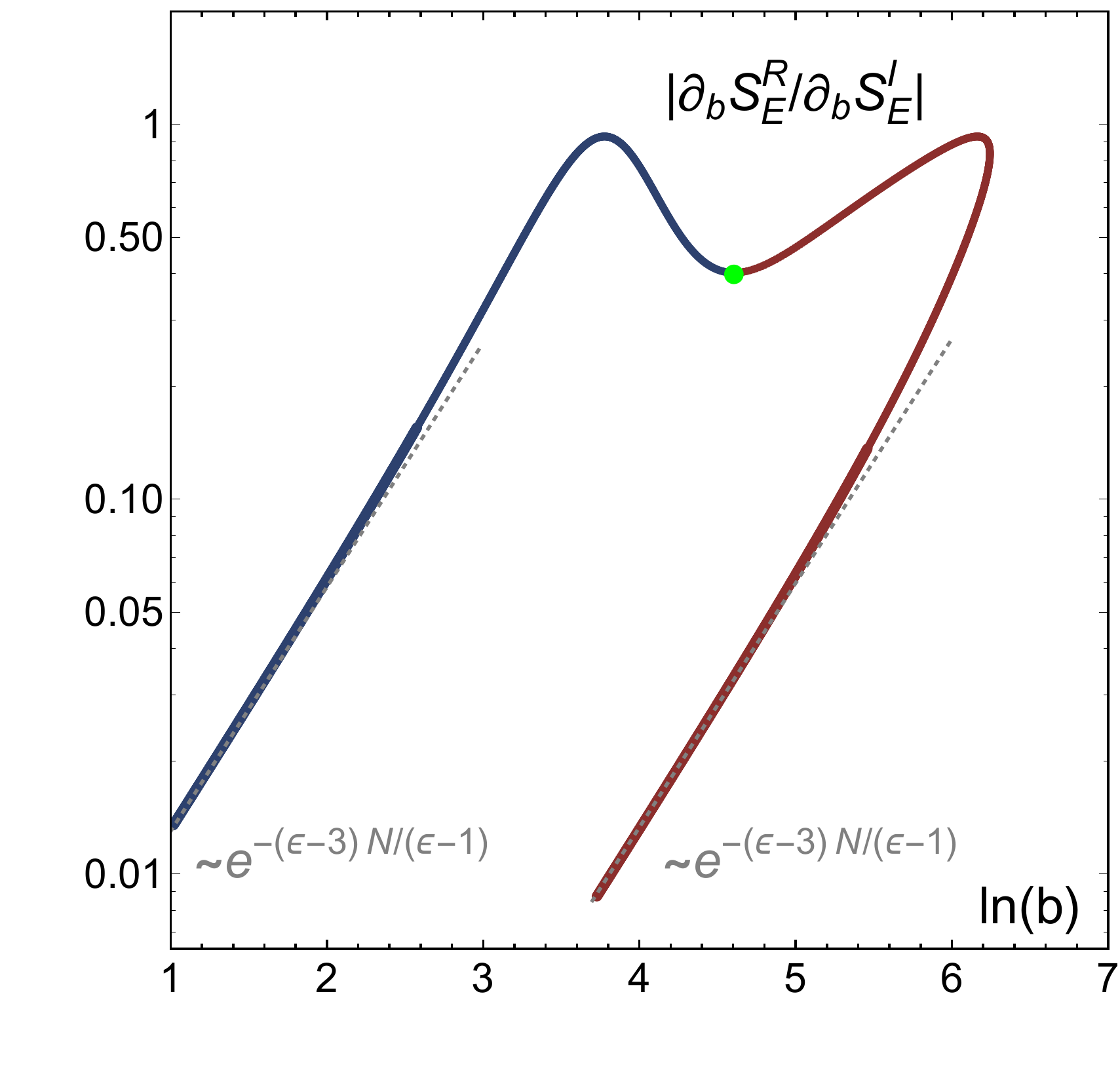}
\end{minipage}%
\begin{minipage}{\smallWidthRight}
\includegraphics[width=\smallWidthRight]{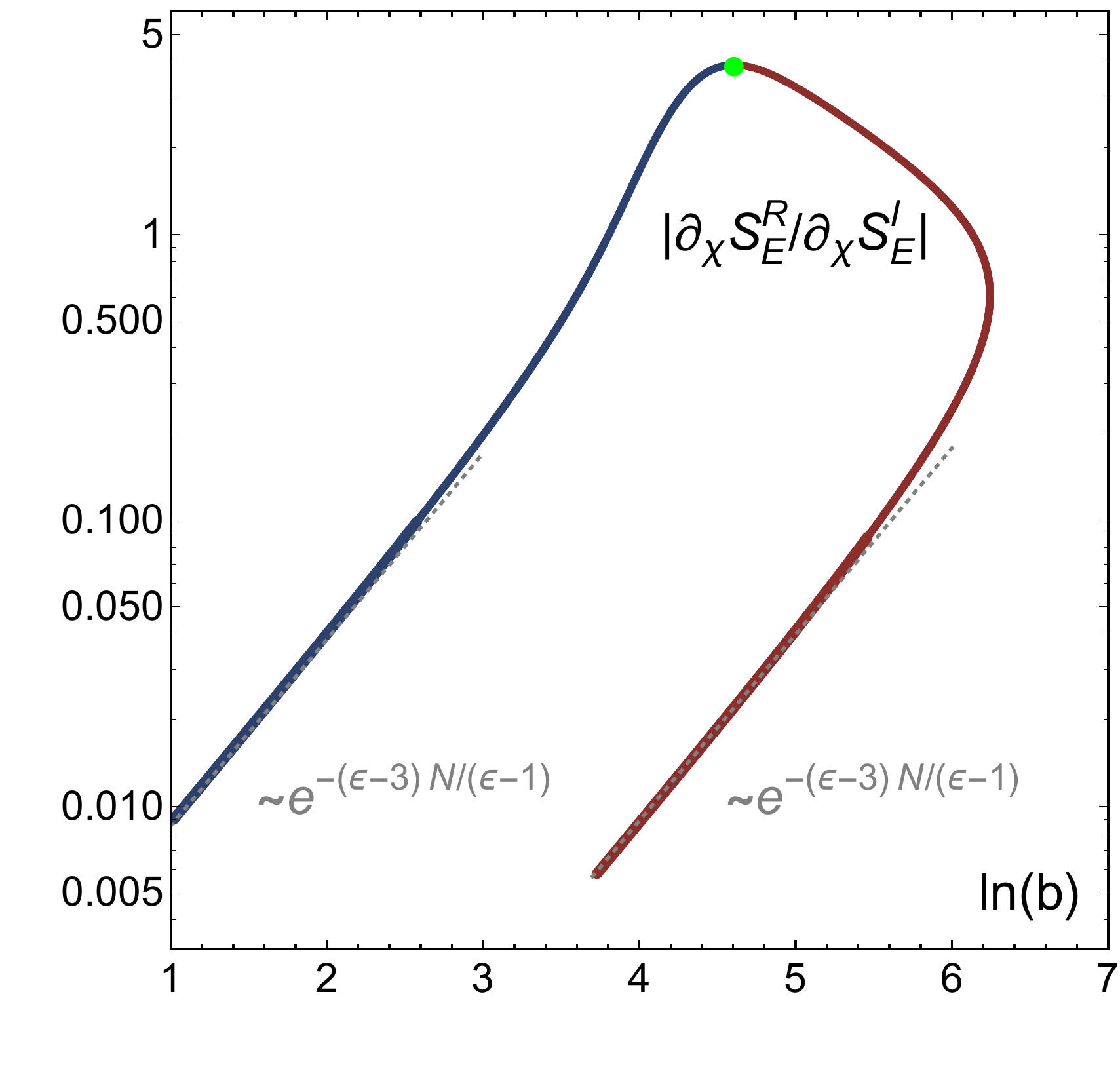}
\end{minipage}%
\caption{ \label{fig:CycWKB} For both histories of interest, the wavefunction becomes classical as the field rolls down the steep negative part of the potential, according to the expected ekpyrotic scaling. However, the initially expanding history is not only likelier, but it also reaches a higher degree of classicality compared to the initially contracting history at a given value of the scale factor.} 
\end{figure}

The main point of interest here is the approach to classicality. The evolution of the WKB conditions \eqref{WKBb} and \eqref{WKBchi} are shown in Fig. \ref{fig:CycWKB}. While the field is on the dark energy plateau, the wavefunction does not become classical: even though condition \eqref{WKBchi} starts being increasingly satisfied (right panel), the same is not true for condition $\eqref{WKBb}$ (left panel). However, as the field rolls down the steep ekpyrotic part of the potential, and as the universe simultaneously contracts, both conditions are increasingly well satisfied, once more according to the scaling relation $\sim e^{-(\epsilon - 3)N/(\epsilon - 1)},$ as expected. Note that the initially expanding history is not only much more likely, it is also much more classical than the always contracting history at a given value $b$ of the scale factor. The curvature scale of the instantons is given by the height of the dark energy plateau \cite{Battarra:2014kga}. Consequently, the involved curvatures are very small (in Planck units), and the semi-classical approximation employed here should be trustworthy. Thus these results show that the no-boundary proposal constitutes a viable theory of initial conditions for the cyclic universe.

\section{Transient Ekpyrosis: Potentials of the Form $V = - \phi^n$} \label{section:phi4}

\begin{figure}[htbp]
\begin{minipage}{\smallWidthLeft}
\includegraphics[width=\smallWidthRight]{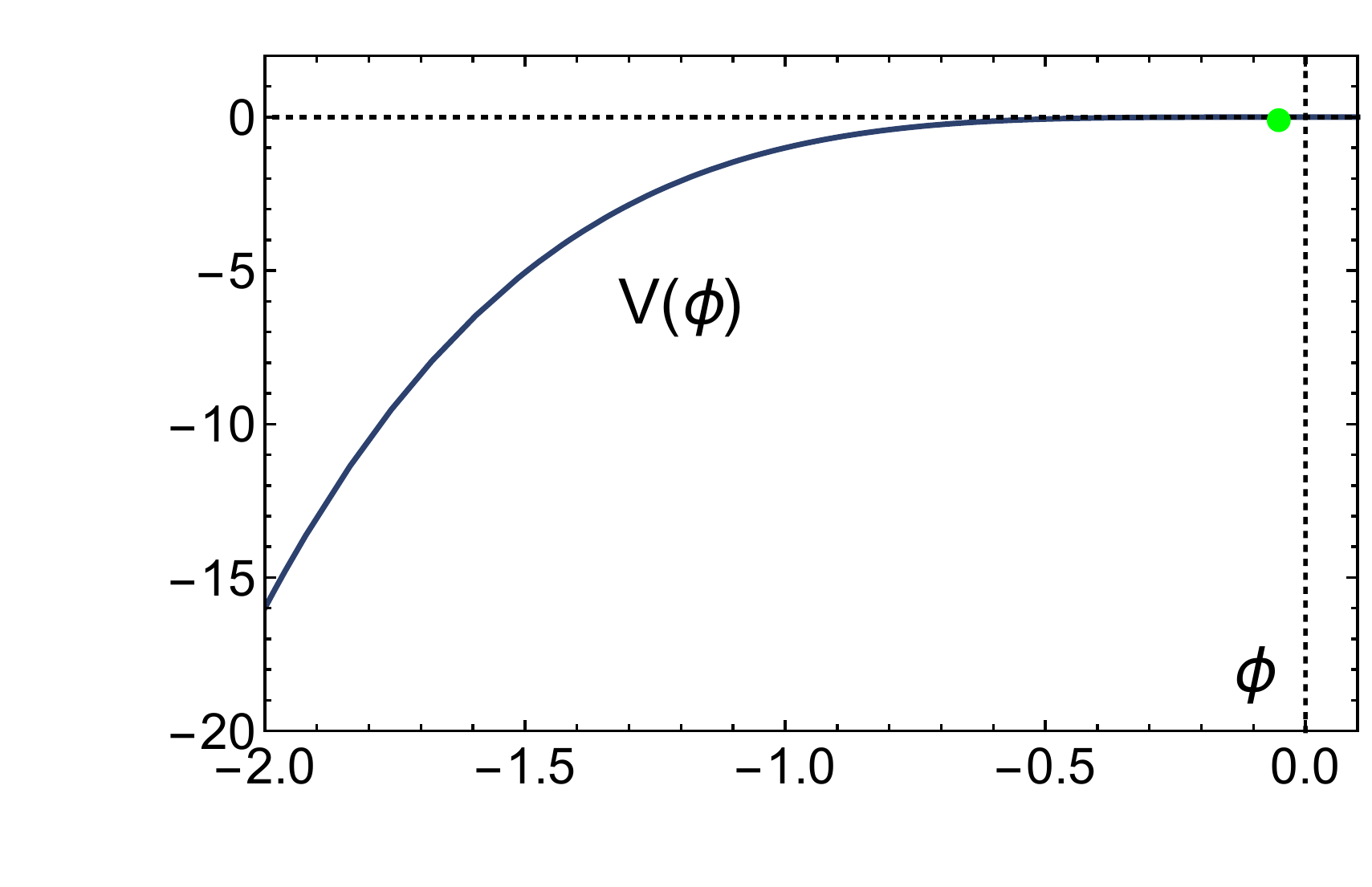}
\end{minipage}%
\begin{minipage}{\smallWidthRight}
\includegraphics[width=\smallWidthRight]{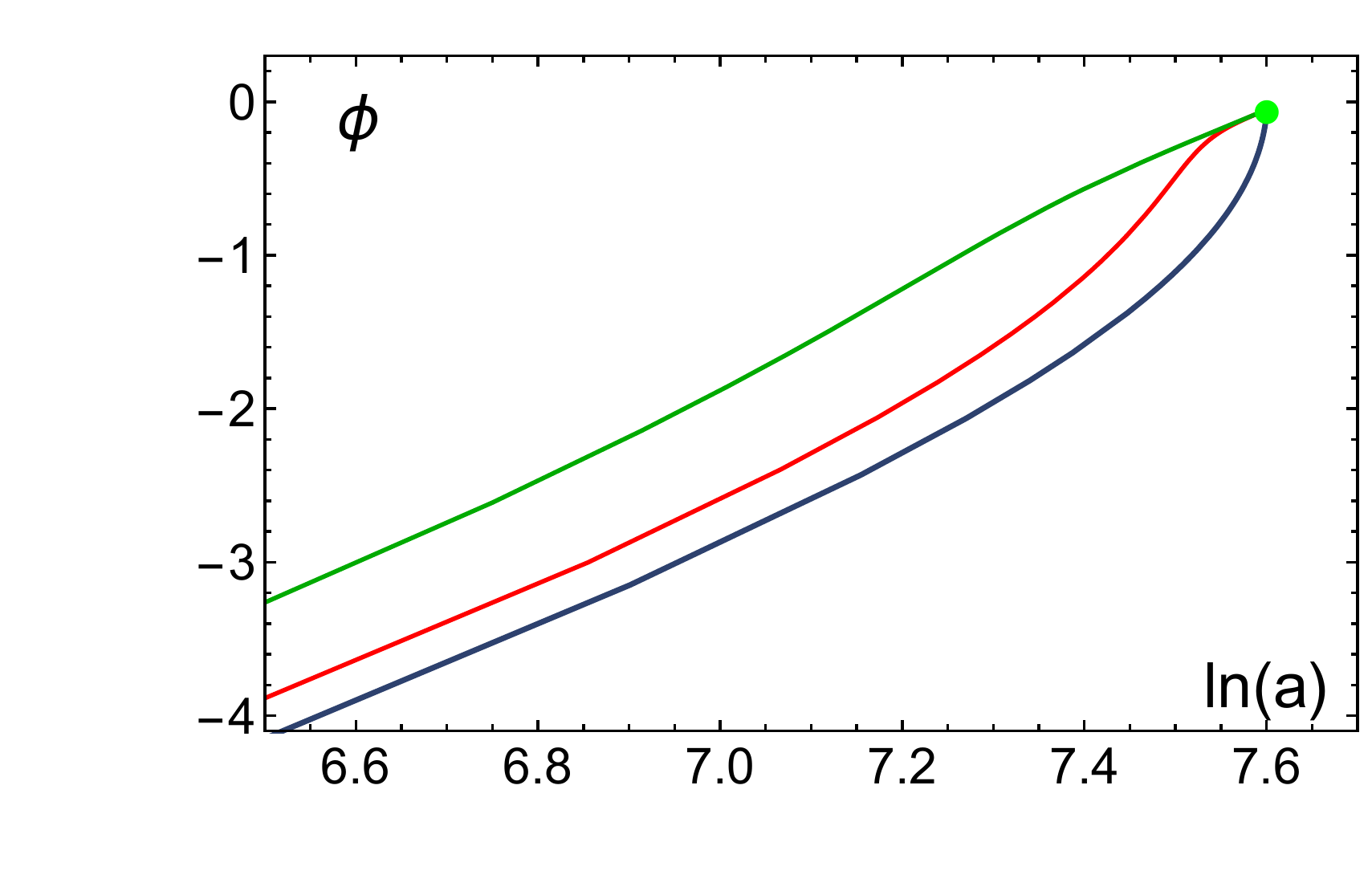}
\end{minipage}%
\caption{ \label{fig:Phi4Pot} Left panel: the potential $V(\phi) = -\phi^4,$ with the initial value $\phi=-1/5$ indicated by the green dot. Right panel: the $3$ histories that we are comparing for this potential. They differ by having increasingly large initial velocities $|\dot\phi|,$ with the colour assignments $\dot\phi_{initial} = -3/800$ (blue), $\dot\phi_{initial} = -3/80$ (red) and $\dot\phi_{initial} = -3/8$ (green).} 
\end{figure}

We have seen that an ekpyrotic phase is both an efficient means of rendering the wavefunction classical, and that it also leads to high (relative) probabilities. For the case where the potential is an exponential function of the field, and where the equation of state is constant, the classicalisation process is efficient and sustained - the final level of classicality reached depends foremost on the total number of e-folds of ekpyrotic contraction. In a general (``landscape'') potential, one may expect that there will be other (sufficiently steep) negative regions of the potential, not of exponential form. For this reason it is interesting to also look at power-law potentials, and here we will consider the potentials $V = - \phi^n$ for $n=4,6,8.$ A further motivation for such potentials stems from the fact that they are used in certain early-universe models, in particular the conformal rolling scenario \cite{Rubakov:2009np} and the pseudo-conformal universe \cite{Hinterbichler:2011qk}. These scenarios effectively use the ekpyrotic mechanism to smoothen the universe, and hence one may ask whether such potentials also provide an efficient means of rendering the wavefunction classical in a WKB sense.

\begin{figure}[htbp]
\begin{minipage}{\smallWidthLeft}
\includegraphics[width=\smallWidthRight]{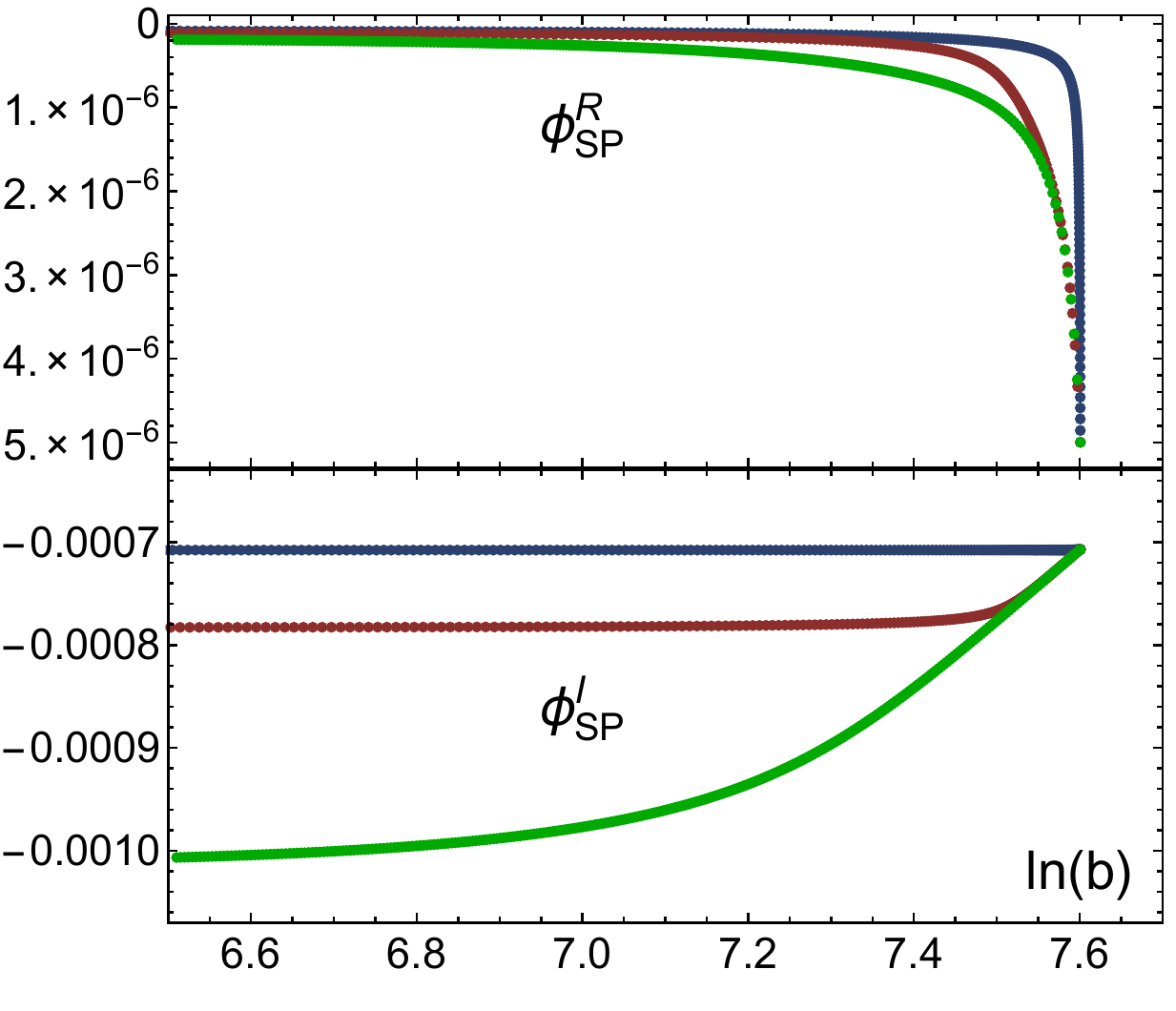}
\end{minipage}%
\begin{minipage}{\smallWidthRight}
\includegraphics[width=\smallWidthRight]{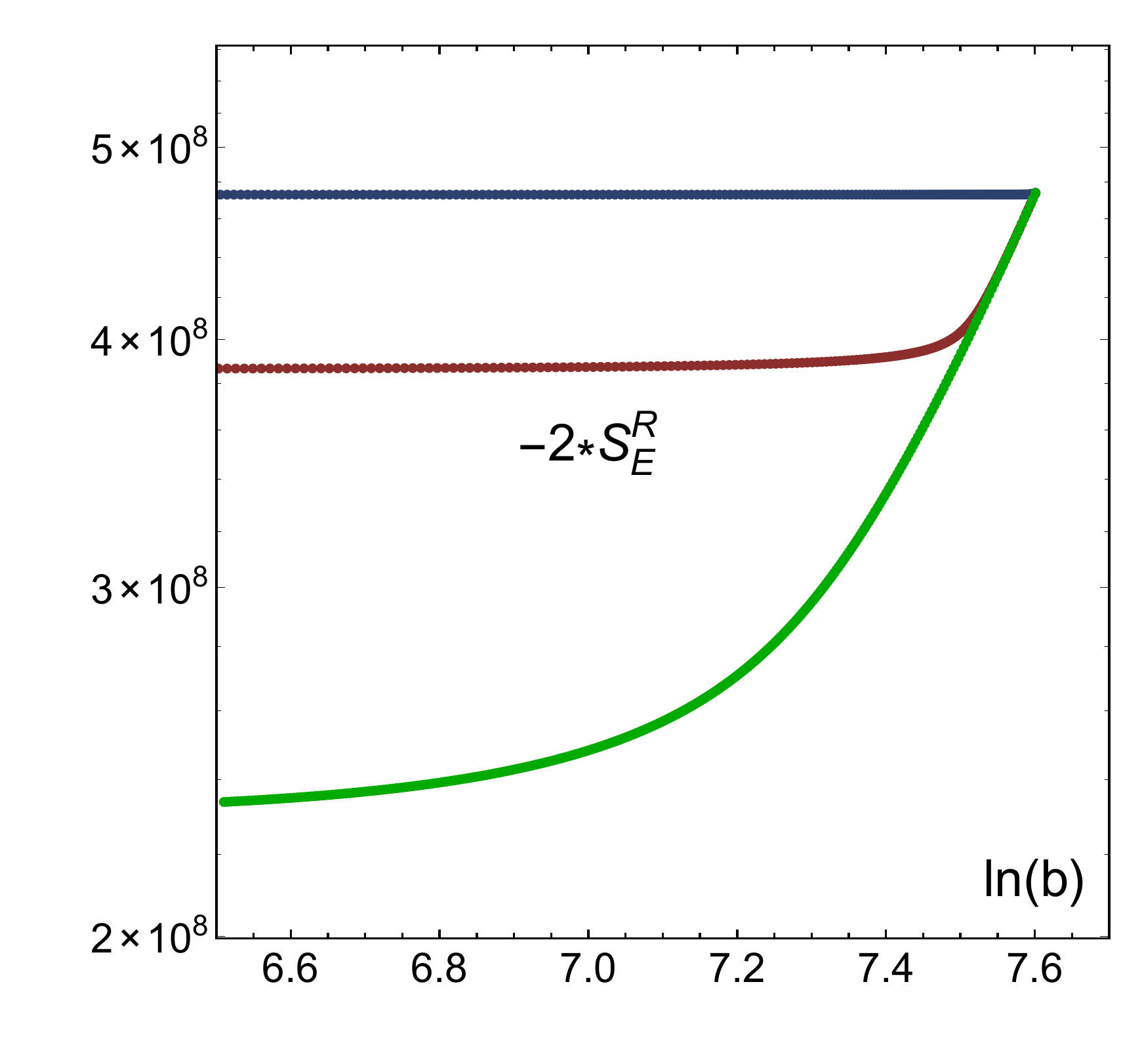}
\end{minipage}%
\caption{ \label{fig:Phi4Phi0ReS} The South Pole values $\phi_{SP}$ and the real part of the Euclidean action $S_E^R$ for the three classical histories shown in Fig. \ref{fig:Phi4Pot}. The colour assignments are the same as in the previous figure.} 
\end{figure}

\begin{figure}[htbp]
\begin{minipage}{\smallWidthLeft}
\includegraphics[width=\smallWidthRight]{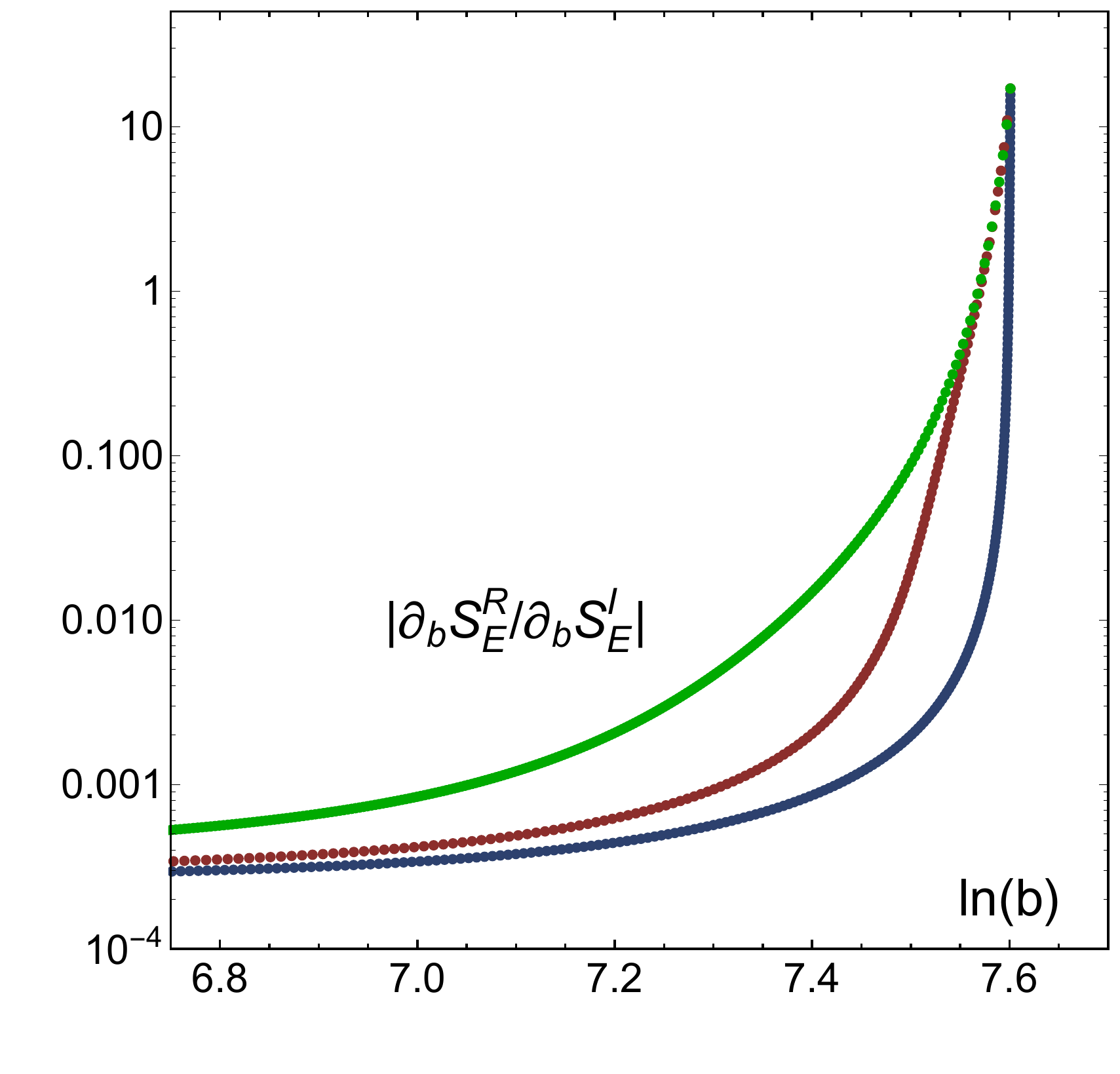}
\end{minipage}%
\begin{minipage}{\smallWidthRight}
\includegraphics[width=\smallWidthRight]{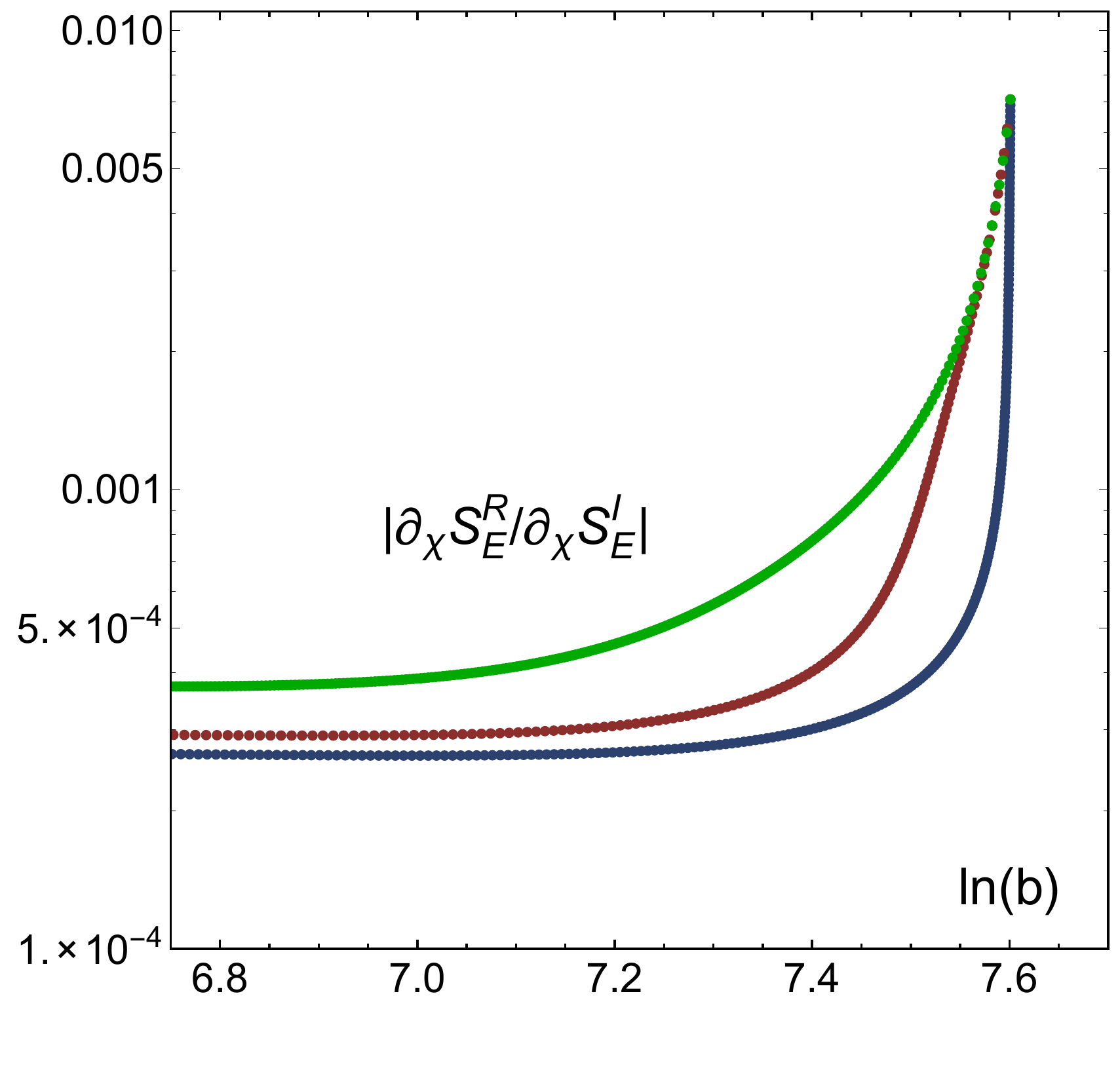}
\end{minipage}%
\caption{ \label{fig:Phi4WKB} The WKB classicality conditions for the three classical histories in Fig. \ref{fig:Phi4Pot}, again with the same colour assignments. After an initial burst of classicalisation, the wavefunction retains the level of classicality attained.} 
\end{figure}

We will analyse two situations: different initial conditions within a given potential, and similar initial conditions but for different potentials. Our treatment is by no means exhaustive, but it is sufficient to reveal a number of interesting effects. We will first fix the potential to be $V = - \phi^4,$ and look at classical histories specified by the following three sets of initial data:
\begin{eqnarray}
a(\lambda_i)  & = 2000, \qquad \dot{a}(\lambda_i) &  = \pm \left( -1 + \frac{a(\lambda_i)^2}{3}(\frac{1}{2} \dot\phi(\lambda_i)^2 + V(\lambda_i))\right)^{1/2} \label{Phi4IC1} \\
\phi(\lambda_i) = & - \frac{1}{20}, \qquad \dot{\phi}(\lambda_i) & = -\frac{3}{800}, -\frac{3}{80}, -\frac{3}{8}\, . \label{Phi4IC2}
\end{eqnarray}
The right panel in Fig. \ref{fig:Phi4Pot} shows the three classical histories following from these initial conditions. The corresponding instantons are specified by South Pole values of the scalar field shown in the left panel of Fig. \ref{fig:Phi4Phi0ReS}, while the right panel of that figure plots the corresponding values of the real part of the Euclidean action. We can observe that classicality is reached surprisingly quickly, especially for the history with the smallest initial value of the scalar field velocity - for this latter history, which is also the one with the highest likelihood, the values of $\phi_{SP}$ and $S_E^R$ stabilise after only a tenth of an e-fold of contraction! Based on this observation, one may expect that the wavefunctions will be highly classical in a WKB sense. The relevant results are plotted in Fig. \ref{fig:Phi4WKB}. Here we see something interesting: while it is true that initially the WKB conditions become increasingly satisfied at a fast rate, they also reach a halt very quickly. This may be understood heuristically as follows: in a steep power law potential, the evolution is dominated by the potential only very early on. During this initial period, the equation of state is typically very large, $\epsilon \gg 1.$ Rather quickly though, the kinetic energy of the scalar field takes over, and one effectively reaches a kinetic dominated phase with $\epsilon \approx 3$. There, in agreement with the results found for a kinetic phase following ekpyrosis in section \ref{section:ekkin}, one sees once more that the process of classicalisation stops. The history with the smallest initial scalar field velocity is also the one that becomes the most classical. Note that the final numerical values for the WKB conditions are pretty small (they are at a level of $10^{-3.5}$) - an interesting question for future work might be to see if nevertheless there might be any effects caused by the remaining traces of ``quantumness''.

\begin{figure}[htbp]
\centering
\begin{minipage}{\fullWidth}
\includegraphics[width=\fullWidth]{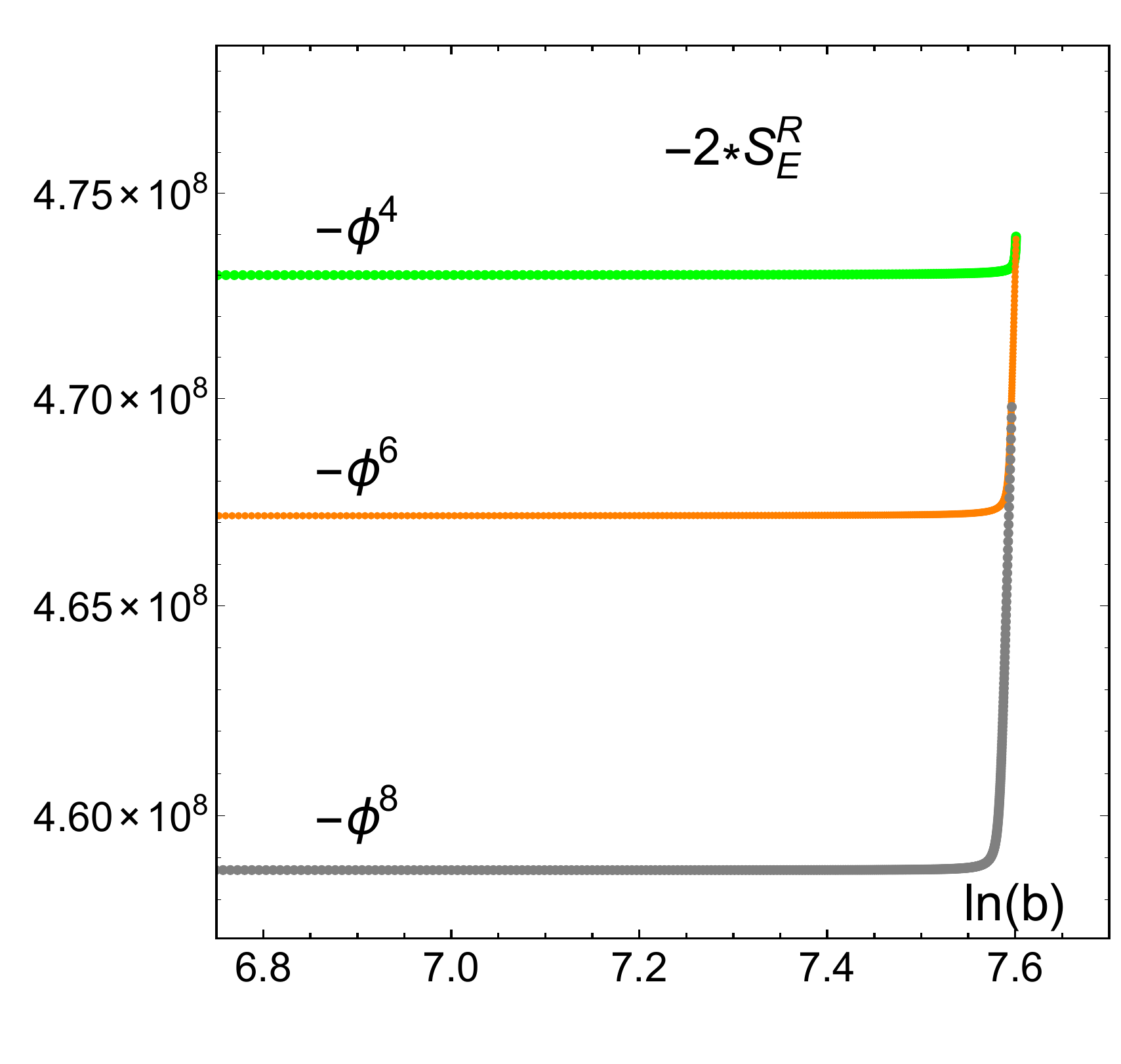}
\end{minipage}%
\caption{\label{fig:PhinAction} A comparison of the relative likelihood of classical histories with small scalar field velocities in the potentials $V = - \phi^{4,6,8}.$ The shallower potential comes out as preferred.}
\end{figure}

A final situation of interest is to compare different potentials. For this purpose, we may analyse the power-law potentials $V = -\phi^{4,6,8}.$ The initial conditions for the classical solutions under consideration are fixed such that the initial values of $a,$ $H$ and $\phi$ coincide with the ones in \eqref{Phi4IC1} - \eqref{Phi4IC2} for the case that $\dot\phi(\lambda_i)=-3/800,$ but with the initial values for $\dot\phi$ adjusted in accord with the Friedmann equation for the potentials $V=-\phi^{6,8}.$ The instantons in all three cases have very similar properties, and they lead to the relative probabilities shown in Fig. \ref{fig:PhinAction}. As can be seen, all three histories quickly lead to a well-defined notion of probability, and the potential that is the least steep ($-\phi^4$) comes out as preferred. 

\begin{figure}[htbp]
\begin{minipage}{\smallWidthLeft}
\includegraphics[width=\smallWidthRight]{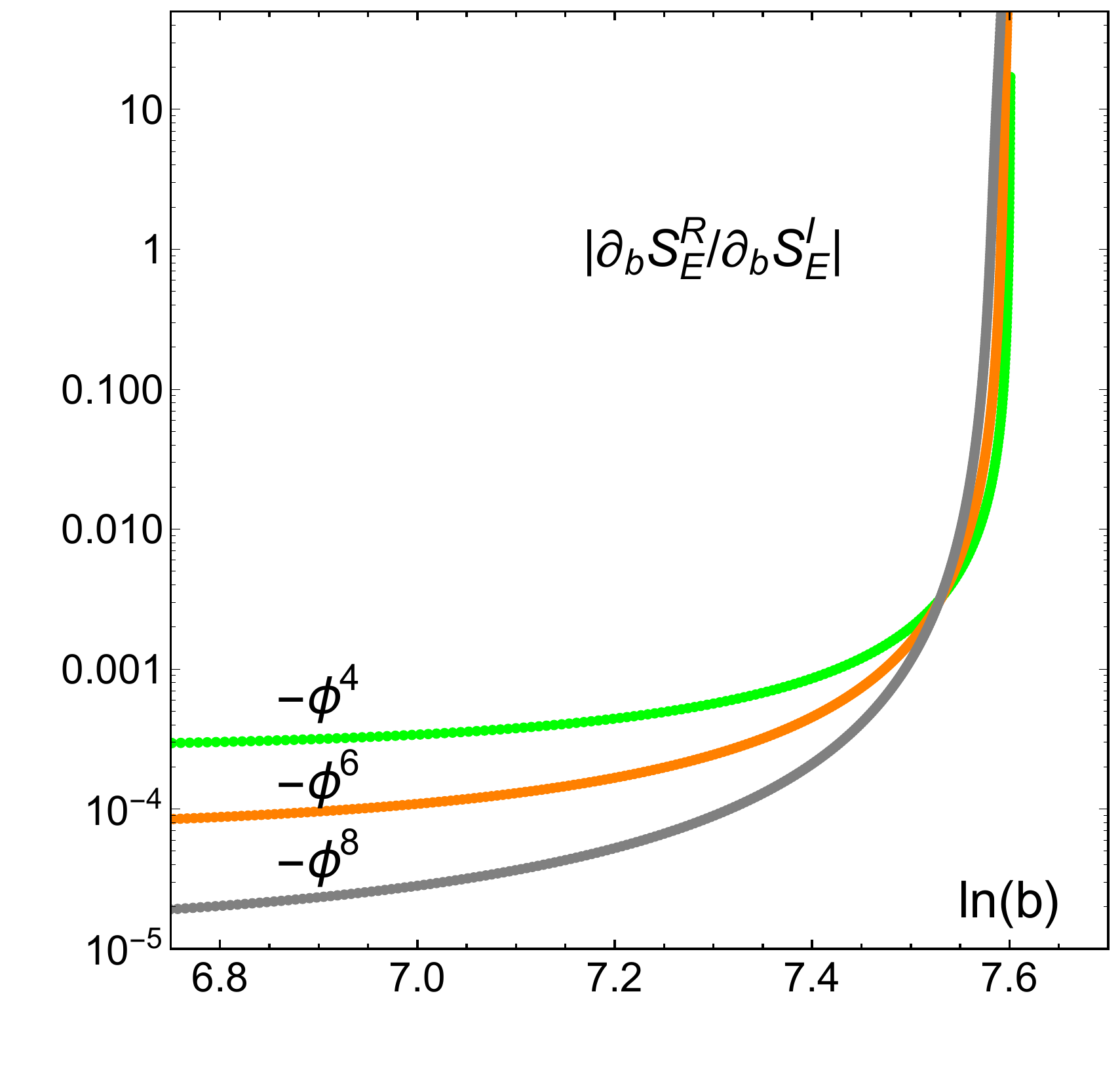}
\end{minipage}%
\begin{minipage}{\smallWidthRight}
\includegraphics[width=\smallWidthRight]{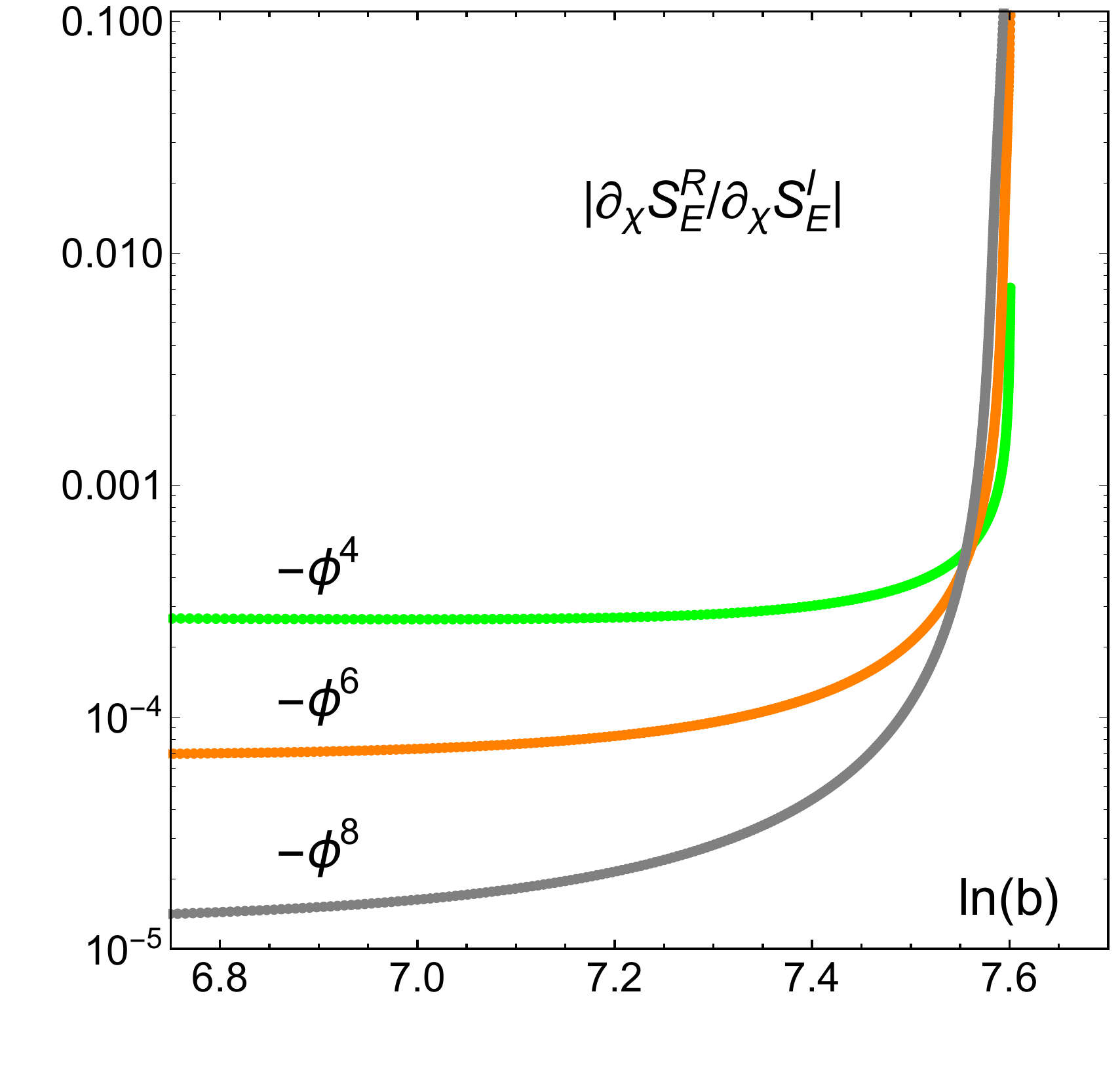}
\end{minipage}%
\caption{ \label{fig:PhinWKB} For potentials of the form $-\phi^n,$ there is an initial burst of classicalisation as the universe contracts. However, after a small amount of contraction, the classicalisation process already stops, as the universe gradually becomes dominated by the kinetic energy of the scalar field.} 
\end{figure}

We can then also take a look at the corresponding WKB classicality conditions, shown in Fig. \ref{fig:PhinWKB}. Here one notices a repetition of the pattern observed for $V=-\phi^4:$ in all three cases, there is an initial burst of classicalisation, which however comes to a halt after just a fraction of an e-fold of contraction. The final level of classicality reached is similar for the three potentials, though the steeper the potential, the more classical the corresponding wavefunction. We have not found an analytic explanation of the final level of classicality attained - this may be an interesting question to pursue in future research. In conclusion, one may summarise the situation by saying that one has a trade-off between likelihood and classicality: steeper power-law potentials lead to universes that are more classical, while shallower potentials come out as likelier.

\section{Discussion} \label{section:Discussion}

The currently most advanced dynamical theory, string theory, suggests that effective potentials may take a very complicated form, with numerous positive, negative, steep and flat regions. In such a situation, especially in a cosmological context, one needs to know how the evolution starts in order to make pre/post-dictions. In the present paper we have investigated the idea, due to Hartle and Hawking, that one may be able to formulate a theory of initial conditions in semi-classical quantum gravity. Their specific proposal is that in the path integral formulation of quantum gravity, one only sums over (complex) 4-geometries that have no boundary to the past. One consequence of this proposal is that the relative probability for various histories of the universe is proportional to a factor $e^{1/|V(\phi_{SP}^R)|},$ where $\phi_{SP}$ is the value of the scalar field at the very bottom of the instanton. This implies that all histories that start out at a small positive or negative value of the potential tend to be preferred\footnote{This is nicely self-consistent with the assumption that a semi-classical approach is valid: for small values of the potential, the Hubble rates and curvature invariants are all small.}. In this way, if the potential is such that it allows for both inflationary and ekpyrotic/cyclic solutions, then ekpyrotic universes (which can start at small negative potential) and cyclic universes (which can start at small positive potential) come out as preferred over cosmologies with large initial expansion or contraction rates, such as inflation\footnote{The existence of such potentials has not been demonstrated conclusively in string theory yet, and thus one must treat these probabilistic statements with a grain of salt.}. Note however that one must be able to join the small initial Hubble rates to the rather high Hubble rates of the early (radiation dominated) hot big bang phase in order to obtain successful model. This can be achieved by having a bounce. A full understanding of bounces is still lacking, but see e.g. \cite{Buchbinder:2007ad,Creminelli:2007aq,Easson:2011zy,Cai:2012va,Koehn:2013upa,Battarra:2014tga,Battefeld:2014uga,Qiu:2015nha} for recent progress. It will be crucial to incorporate into the currently known ekpyrotic instantons a (non-singular) bounce in order to obtain a full semi-classical history of the universe. Results regarding this issue will be presented in a forthcoming publication.

In addition to figuring out what a likely starting point for cosmology might look like, one would also like to find an explanation for why spacetime and matter came to behave so classically, even in the very early universe. In calculating predictions of cosmological models, one typically quantises small quantum fluctuations around a classical background - but why is it justified to assume a classical background in the first place? Here we have seen that the no-boundary proposal provides such an explanation. In fact, we were able to derive in a rather precise fashion how fast a classical spacetime description becomes meaningful, by calculating the WKB classicality conditions for a range of theories and potentials. For exponential potentials, both positive and negative, we have found that the classicality conditions are satisfied in proportion to a factor 
\begin{equation}
WKB \, \propto \, e^{-\frac{\epsilon - 3}{\epsilon - 1}N},
\end{equation}
where $N$ denotes the number of e-folds of evolution. Since this factor must approach zero, we can see that this scaling singles out two possible regimes: $\epsilon 
< 1,$ corresponding to inflation, or $\epsilon > 3,$ corresponding to ekpyrosis. Thus, we can see that the only two dynamical smoothing mechanisms for the universe share the very fundamental property of also rendering spacetime and matter classical! This conclusion is also supported by the Wheeler-deWitt equation, as shown in section \ref{section:QC}, and it lends strong support to the idea that (at least one of) these two types of theories must have played a crucial role in the early universe. Note also the implications for cosmological scenarios that do not incorporate either an inflationary or an ekpyrotic phase: they must rely on strong (and unjustified) assumptions about the classicality and configuration of spacetime and matter at very early times. An additional aspect that we have probed in the present paper is the issue of what happens when the ekpyrotic phase comes to an end, and the universe becomes kinetic dominated. Here we have found that the level of classicality reached up to that point remains preserved during the kinetic phase (when $\epsilon = 3$). This is important, especially for non-singular bounce models, as it means that the universe approaches the bounce in a highly classical state. 

We have also analysed negative potentials of power-law form. For these potentials, the dynamics is at first dominated by the potential, and the equation of state is very large. Correspondingly, there is an initial burst of classicalisation, during which the WKB conditions become satisfied very rapidly. Then, as the evolution becomes dominated by the kinetic energy of the scalar field, the equation of state drops down to $\epsilon = 3$ and the level of classicality remains constant. It would be interesting to understand what determines the final level of classicality achieved.  

It may be interesting to finish with a few remarks regarding the arrow of time. As is well-known, time does not appear explicitly in quantum cosmology, and so, for instance, one may ask why a given ekpyrotic history is contracting and becoming classical rather than expanding and becoming less classical. In minisuperspace there is in fact no good way to tell. However, one may easily imagine extensions of the present framework where a second field and its fluctuations are included. Then, on top of having an increasingly classical background, small quantum (entropy) fluctuations also become amplified. These fluctuations evolve into a highly squeezed state, which may be re-interpreted as a statistical ensemble of classical fluctuations. What is more, when the entropy perturbations get converted to curvature perturbations, decoherence occurs very efficiently \cite{Battarra:2013cha}. These processes thus provide an unambiguous arrow of time - the fluctuations provide a time direction. Note that for a single-field ekpyrotic phase, such an arrow of time does not arise: in that situation the quantum fluctuations are not amplified, and the universe remains completely empty. Thus, in the ekpyrotic case, the first scalar causes spacetime to become classical, and a second scalar plays the additional role of providing an arrow of time.


\acknowledgments

I would like to thank Rhiannon Gwyn, Claus Kiefer, George Lavrelashvili and Paul Steinhardt for stimulating discussions and correspondence. I gratefully acknowledge the support of the European Research Council via the Starting Grant Nr. 256994 ``StringCosmOS''.

\bibliographystyle{utphys}
\bibliography{InflEkpUniverses}

\end{document}